\documentclass[twocolumn,runningheads]{svjour2}

\smartqed  

\usepackage{graphicx}
\usepackage{longtable}
\usepackage{color}

\usepackage{multirow}
\usepackage{supertabular}

\begin{document}

\journalname{International Journal of Advanced Manufacturing Technology }

\title{Displacements analysis of self-excited vibrations in turning. }


\author{Claudiu F. Bisu \and Philippe~Darnis \and Alain~G\'erard \and Jean-Yves~K'nevez}

\institute{C.F. Bisu \at University Politehnica from Bucharest,\\313 Splaiul Independentei,  060042 Bucharest Roumanie (UE)
              \\\email{cfbisu@gmail.com}           
           \and Claudiu F. Bisu \and Alain~G\'erard (corresponding author) \and Jean-Yves~K'nevez \at
              Universit\'e de Bordeaux \\351 cours de la Lib\'eration, 33405 Talence-cedex France (UE)\\
              Tel.: +33 (0)5 40 00 62 23\\
              Fax: +33 (0)5 40 00 69 64
              \\\email{alain.gerard@u-bordeaux1.fr}   
\and P. Darnis \at Universit\'e de Bordeaux - IUT EA 496,\\15 rue Naudet, 33175 Gradignan Cedex France (UE) \\\email{philippe.darnis@u-bordeaux1.fr} }
\date{Received:  / Accepted: }

\maketitle

\begin{abstract}
The actual research deals with determining by a new protocol the necessary parameters considering a three-dimensional model to simulate in a realistic way the turning process on machine tool. This paper is dedicated to the experimental displacements analysis of the block tool / block workpiece with self-excited vibrations. In connexion with turning process, the self-excited vibrations domain is obtained starting from spectra of two accelerometers. One three axes accelerometer placed on the tool and one unidirectional accelerometer placed on the front of bearing of spindle. The existence of a displacements plane attached to the tool edge point is revealed. This plane proves to be inclined compared to the machines tool axes. This plane contains an ellipse that is the place of the points of the tool tip displacements. We establish that the tool tip point describes an ellipse. This ellipse is very small and can be considered as a small straight line segment for the stable cutting process (without vibrations). In unstable mode (with vibrations) the ellipse of displacements is really more visible. A difference in phase occurs between the tool tip displacements on the radial direction and on the cutting one. The feed motion direction and the cutting one are almost in phase. The values of the long and small ellipse axes (and  their ratio) shows that these sizes are increasing with the feed rate value. A weak growth (6\%) of the long and small axes ratio is obtained when the feed rate value decreases. The axis that goes through the stiffness center and the tool tip represents the maximum stiffness direction. The maximum (resp. minimum) stiffness axis of the tool is perpendicular to the large (resp. small) ellipse displacements axis. The  self-excited vibrations appearance is strongly influenced by the system stiffness values, their ration and their direction. FFT analysis of the accelerometers signals allows to reach several important parameters and establish coherent correlations between tool tip displacements and the static - elastic characteristics of the machine tool components tested (see \cite{bisu-07}).


\end{abstract}

\section*{Nomenclature}
\label{sec:Nomenclature}

	\begin{tabular}{lp{6,8cm}}  
\raggedright 

$ap$ & Depth of cut (mm)\\
$a_{u}$ ($b_{u}$ ) & Large (small) axis of ellipse attached with the  cutting tool tip displacements (m)\\
$a_{f}$ ($b_{f}$)& Large (small) axis of ellipse attached with the forces charge points (m)\\
\textbf{BT} & Block Tool\\
\textbf{BW} & Block Workpiece\\
$f$  & Feed rate (mm/rev)\\
$N$ & Number of revolutions (rpm)\\
$n_{f}$ & Vector attached to the components system stiffness \\
$n_{u}$ & Normal direction of the plane $P_{u}$\\
$n_{ui}$ & Unitary vector, support of the ellipse axis i ($i= a,b$) situated in the plane $P_{u}$ \\
$P_{u}$ & Plane attached with the cutting tool edge displacements\\
$r_{\epsilon}$ & Cutting edge radius (mm)\\
R & Cutting tool edge radius (mm) \\
$R_{t}$ & Total roughness $(\mu m)$\\
	\end{tabular} 
	
\section*{Nomenclature}
\label{sec:Nomenclature}

	\begin{tabular}{lp{6,8cm}}
$T$ & Time that corresponding to one workpiece revolution \\
u & Tool edge point displacement (m)\\	
x (z) & Radial (axial) direction \\
y & Cutting feed axis\\
$\alpha$ & Clearance angle (degree)\\
$\alpha_{\kappa(xy)}$ & Angle of the main stiffness direction in the plane (x,y) (degree)\\
$\alpha_{\kappa(yz)}$ & Angle of the main stiffness direction in the plane (y,z) (degree)\\
$\Delta t$ & Temporal dephasing between two signals\\
$\varphi_{u}$ & Dephasing between the displacements components attached to the tool edge point (degree)\\
$\gamma$ & Cutting angle (degree)\\
$\lambda_{s}$ & Inclination angle of edge (degree)\\
$\kappa_{r}$ & Direct angle (degree)\\
$\theta_{e(xy)}$ & Main displacements direction angle of the tool edge point in the plane (x,y) \\
$\theta_{e(yz)}$ & Main displacements direction angle of the tool edge point in the plane (y,z) \\

	\end{tabular}

\section{Introduction}
\label{sec:1}

The machine tools dynamic phenomena come from the elastic system machine / cutting process interaction  \cite{cowley-70}, \cite{deacu-pavel-77}, \cite{dimla-04}, \cite{ispas-AA-anghel-99}. This interaction is thus the dynamic aspects generating source classically met in the machine tools use \cite{dassanayake-suh-08}, \cite{kegg-65}. The cutting process actions applied to the elastic machining system cause relative tool / workpiece  displacements that generate vibrations \cite{deciu-A-dragomirescu-02}. These modify the chip section, the contact pressure, the speed of relative movement etc. Consequently, the cutting process instability can cause the dynamic system instability of the machine tool: vibrations appear \cite{merrit-65}. They have undesirable effect on the machined workpiece surface quality and on the tool wear \cite{benardos-A-vosniakos-06}, \cite{peigne-03}. Maintenance problems may occur, even ruptures of the machine tools elements \cite{tlusty-polacek-63}. Thus, it is necessary to develop numerical and experimental models ensuring to study the vibratory phenomena during machining, in order to obtain the stable process in cutting \cite{chiou-A-liang-95}, \cite{dimla-04}, \cite{li-01}, \cite{thevenot-05}.

Considerable efforts were developed to model the cutting process correctly. However, at the present time the suggested solutions are still far from providing sufficiently relevant and general models to translate in an acceptable way all the available experimental results \cite{cahuc-AA-battaglia-01}, \cite{cahuc-A-laheurte-07}. Models become very complex for studying the three-dimensional model and nonstationary cutting conditions, particularly in the case of the vibratory situations, \cite{benmohammed-96}, \cite{chen-tsao-06}, \cite{dassanayake-suh-08}, \cite{marot-80}, \cite{mei-05}, \cite{segreti-02}. A model for the expression of cutting forces uncertainty in different cutting conditions was proposed \cite{axinte-A-chiffre-01}. The model enables the uncertainty prediction of cutting force measurements, for a defined cutting parameters range, based on calibration data and measurements collected in a few experiments run using a factorial design. The results showed that the contributions in the uncertainty budget from single cutting force measurements are balanced between the calibration error on the channel where the force is measured, and the cumulative error due to the cutting parameters variation. Being responsible for approximately 50\% of the uncertainty in single force experiments, the cutting parameters variation may be responsible for the scatter of cutting force measurements found in the literature. 

Today, the machine tools have rather high dynamic behaviours with regard to rigidities and damping capacities. However, the causes of the vibrations appearance, or unstable modes, are given by the the cutting dynamics process under various working conditions. In other words, under certain conditions, the cutting interaction process with the machine tool elastic system causes the appearance of vibrations. Self-sustained vibrations are observed. These vibrations, due to the cutting process, have frequencies close to the natural system frequencies, and are generated by the cutting actions variation, variation which depends on the various parameters characteristic of the process \cite{calderon-98}, \cite{calderon-AA-ispas-98}, \cite{deciu-A-dragomirescu-02}.

These parameter  modifications lead to the variations of relative displacements of tool / workpiece and tool / chip. Variations of the chip section are generated, inducing variations of the cutting actions, and thus maintaining the vibrations. It is thus advisable to take into account these parameter variations in a three-dimensional model cutting process which requires a specific determination. These problems were already studied in the past. Different methods were used to estimate workpiece elastic deflections under cutting forces in turning by using analytical modelling of workpiece elastic deflections, by finite element (via Castigliano's second theorem) \cite{benardos-A-vosniakos-06}, and by finite-differences \cite{qiang-00}. In \cite{li-01}, a real-time compensation scheme is developed based on a cutting force induced error model.

In this paper, an experimental study is developed to determine and characterize a series of parameters necessary for the dynamic model of cutting process in agreement with static aspects obtained previously \cite{bisu-07}. After having exposed the tests protocol used (see Sect.~\ref{sec:2}), displacements are measured and a frequency analysis is carried out (see Sect.~\ref{sec:3}). A displacement plane attached of the tool point (see Sect.~\ref{sec:4}) is highlighted starting from the accelerometer data. The deep analysis included in Section~\ref{sec:5} shows that these displacements describe an ellipse. The evolution of the ellipse axis ratio can be characterized by a slightly increasing function depending on the feed motion while their ratio follows an opposite law. This ellipse is located in a tilted precise plane compared to the machine tool axes. Before concluding, Section~\ref{sec:6} gives a correlation between the tool tip stiffness and displacements on the one hand, and on the other hand between stiffness center \cite{bisu-07} and the central axis of the dynamic cutting process.

\section{Test protocol}
\label{sec:2}

\subsection{Approach}
\label{demarche}

In turning operation, the workpiece has a rotary motion whereas the tool has a linear translation. It is obvious that both the tool and workpiece vibrate in material removal. Although workpiece vibrations impact both cutting instability and product quality including surface finishing, most models developed for investigating surface roughness \cite{grzesik-96}, \cite{sahin-motorcu-05}, \cite{thomas-AA-masounave-96} do not consider workpiece vibrations at all.

A new tests protocol is here proposed to determine a series of led phenomenologic parameters during the cutting process in particular in the self-sustained vibrations case. According to \cite{bisu-07}, the machining system static behavior presents couplings in all the directions. It is thus advisable to expect that it is the same in dynamics.

\begin{figure}[htbp]
\centering
		\includegraphics[width=0.48\textwidth]{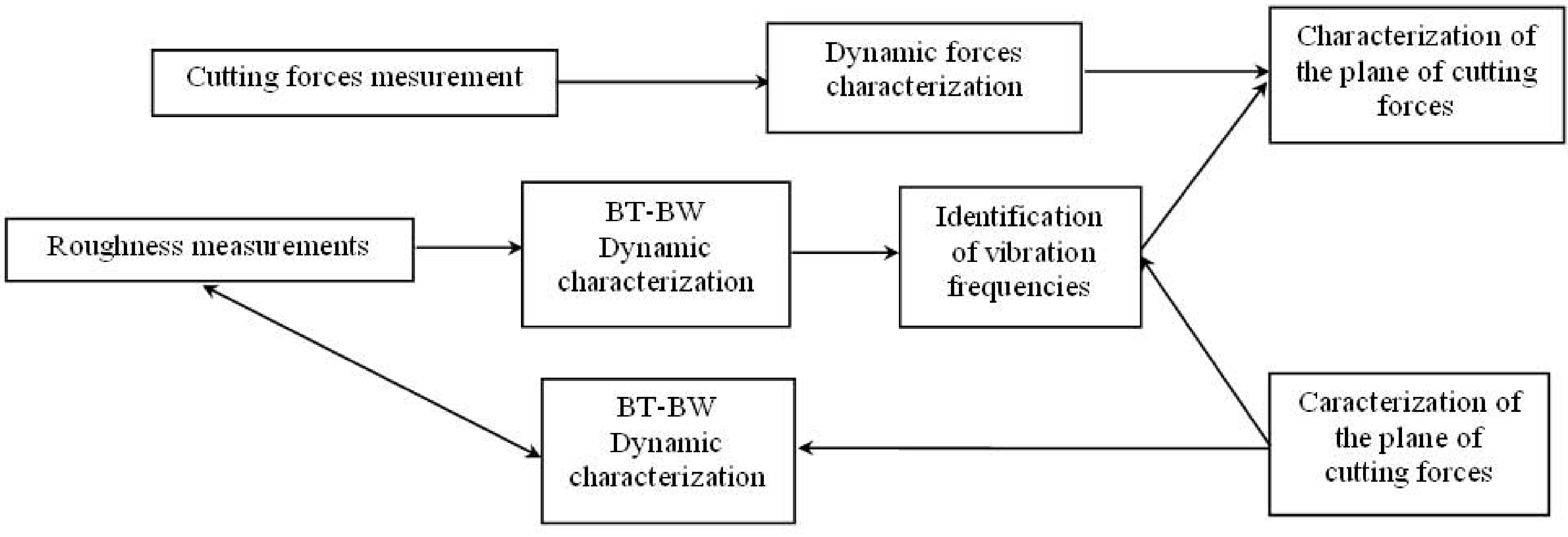}
	\caption{General approach considered}
	\label{fig1}
\end{figure}

The approach adopted for this study constitutes a part of a global vision of research presented in the Fig.~\ref{fig1}. As in the static case, the dynamic characterization is carried out starting from the interaction of two elements: block tool (\textbf{BT}) and block workpiece (\textbf{BW}) forming the machining system and knowing that, a priori, each element taken separately have a different vibratory behaviour (cf. figure~\ref{fig3}), \cite{bisu-AAAAA-ispas-07a}.

\subsection{Experimental device}
\label{dispositif}

\begin{figure}[htbp]
\centering
		\includegraphics[width=0.48\textwidth]{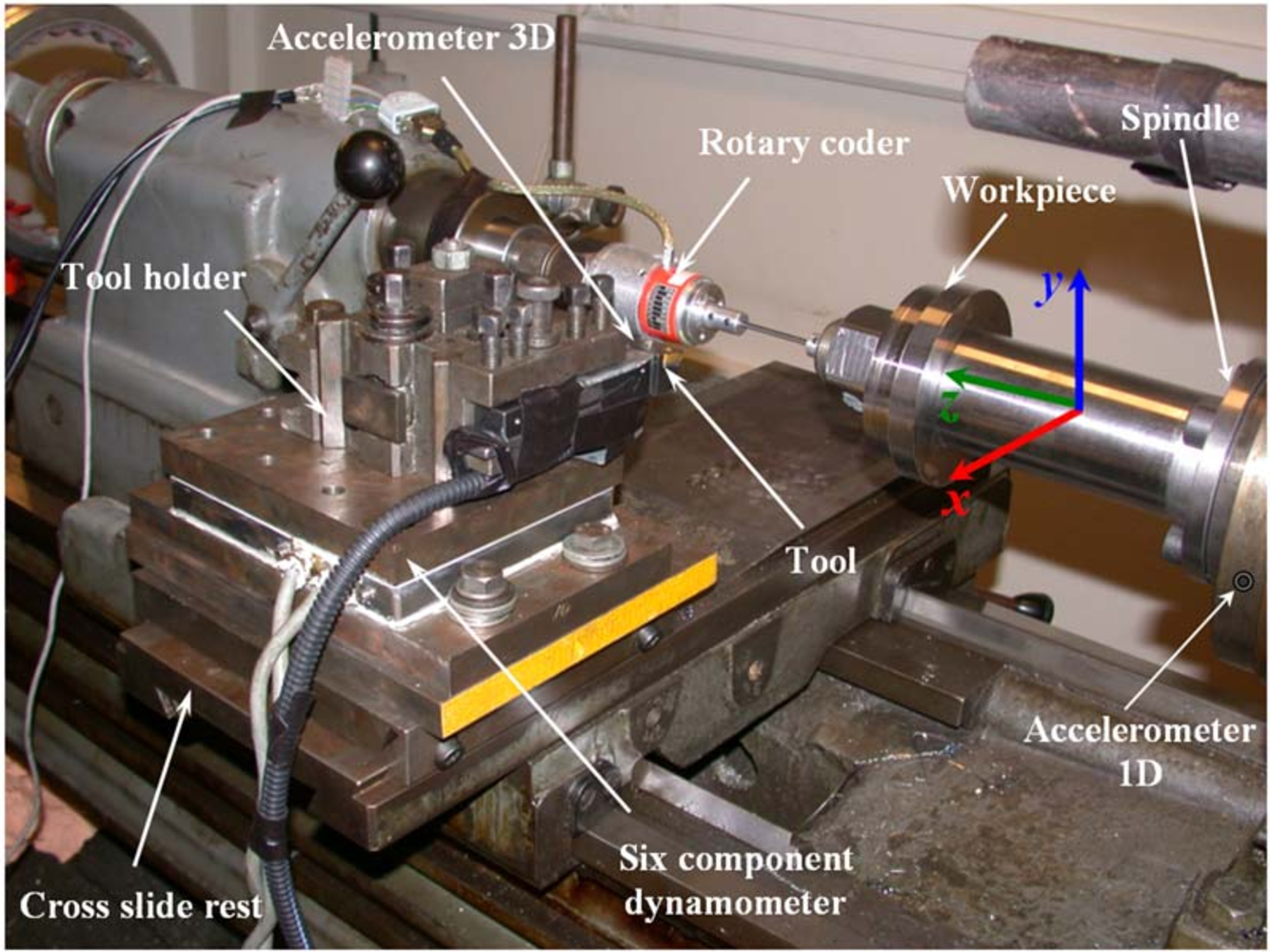}
	\caption{Experimental machining system and metrology environment in turning process}
	\label{fig2}
\end{figure}

Dynamic cutting tests are carried out on a conventional lathe (Ernault HN 400) for which the spindle speed does not exceed 3~500~rpm. The main components of the test machining system used are presented without its data-processing environment in the Fig.~\ref{fig2}. The machining system behavior is analyzed through one three-direction accelerometer fixed on the tool and using two unidirectional accelerometers positioned on the lathe, on the front bearing of the spindle, to identify the influence of this one on the cutting process. Moreover, a six-components dynamometer \cite{couetard-93}, being used as tool-holder \cite{toulouse-98}, is positioned on the lathe to measure all the cutting forces (forces and torques). This study is presented in the paper \cite{bisu-AAAAA-ispas-07a}. The dynamometer used is an evolution of \cite{couetard-93}. The active part made of the strain gauges is replaced by piezoelectric sensors adapted for faster acquisitions and being adequate and compatible with the vibratory phenomena necessary to measure. 

The three-dimensional dynamic character is highligh-ted by seeking the various existing correlations between the various parts of the machining system and the various parameters evolutions, which ensure to characterize the process.

\subsection{Data acquisition protocol}
\label{protocole}

\begin{figure}[htbp]
\centering
		\includegraphics[width=0.48\textwidth]{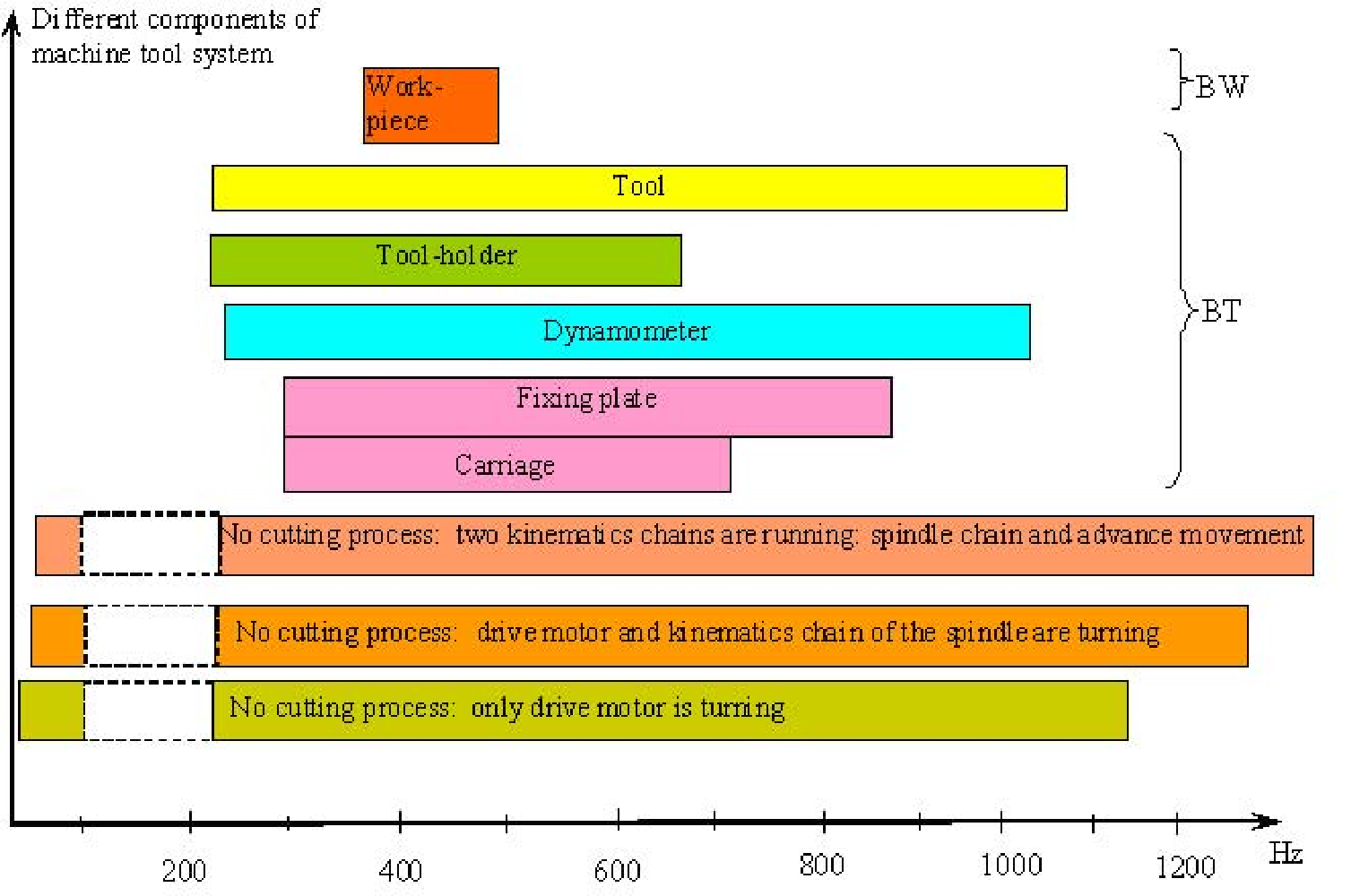}
	\caption{Natural frequencies domain of different components of the machining system}
	\label{fig3}
\end{figure}

The self-excited vibrations characterization method, is based on various work  \cite{kudinov-70}, \cite{moraru-A-rusu-79}, \cite{rusu-75} and on the results of the preliminary dynamic study undertaken in \cite{bisu-AA-knevez-06}, is summarized in Fig.~\ref{fig3}.

The static study undertaken here (cf. \cite{bisu-07} and \cite{bisu-AAAA-ispas-06a}) allows to identify perfectly the machine tool frequency domain and each one of its components under various operating conditions. Moreover, a virgin frequency band at any natural frequency was not found. The machine tool was turned on without part to machine. Thus, the appearance of frequencies in this zone when the cutting process exists, highlights the vibratory phenomenon induced by this one. In what follows, the cutting parameters are selected constant, excluding the depth of cut ($ap$).

On the test machining system, the instantaneous spe-ed of rotation is controlled by a rotary encoder directly related to the workpiece. The connection is carried out by a rigid steel wire, which allows a better transmission of the behavior (Fig.~\ref{fig2}). During the cutting process, the number of revolutions is controlled constantly at nearly 690~rpm and a negligible variation cutting speed of about 1~\% is detected. 

\begin{figure}[htbp]
\centering
		\includegraphics[width=0.45\textwidth]{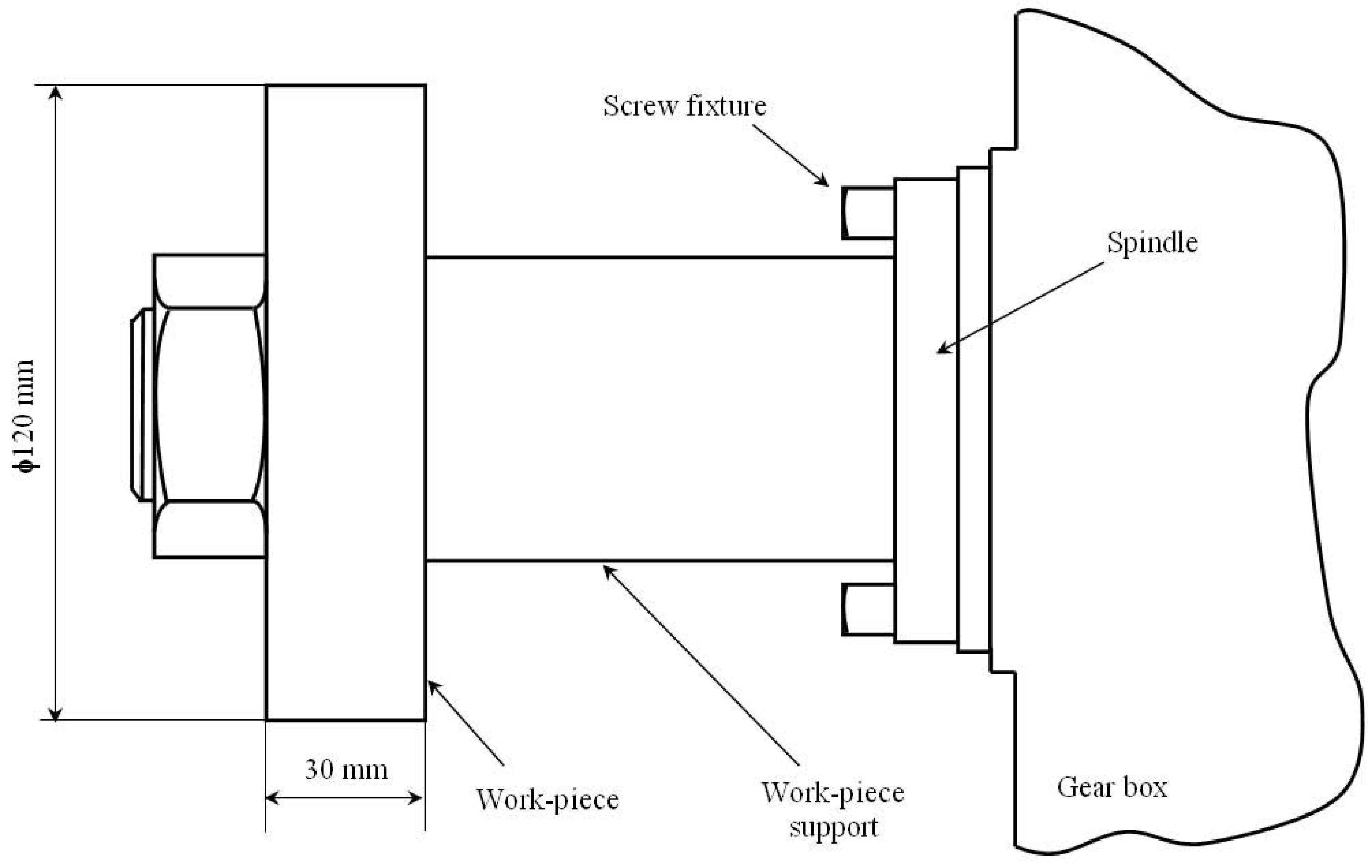}
		\caption{Tested workpiece}
	\label{fig4}
\end{figure}

All measurements are collected on a PC and are thus treated in real time. For these tests, the used tool was a noncoated carbide tool (TNMA~16~04~12) without chip break geometry. The cutting material is a chrome molybdenum alloy type (AiSi 4140). The cylindrical test-tubes have a diameter of 120~mm and a length of 30~mm (Fig.~\ref{fig4}). The dimensions retained for these test-tubes were selected using the finite elements method coupled to an optimization method using SAMCEF$^{\textregistered}$ software presented in \cite{bisu-AAAA-ispas-06a}.

\begin{figure}[htbp]
\centering
		\includegraphics[width=0.35\textwidth]{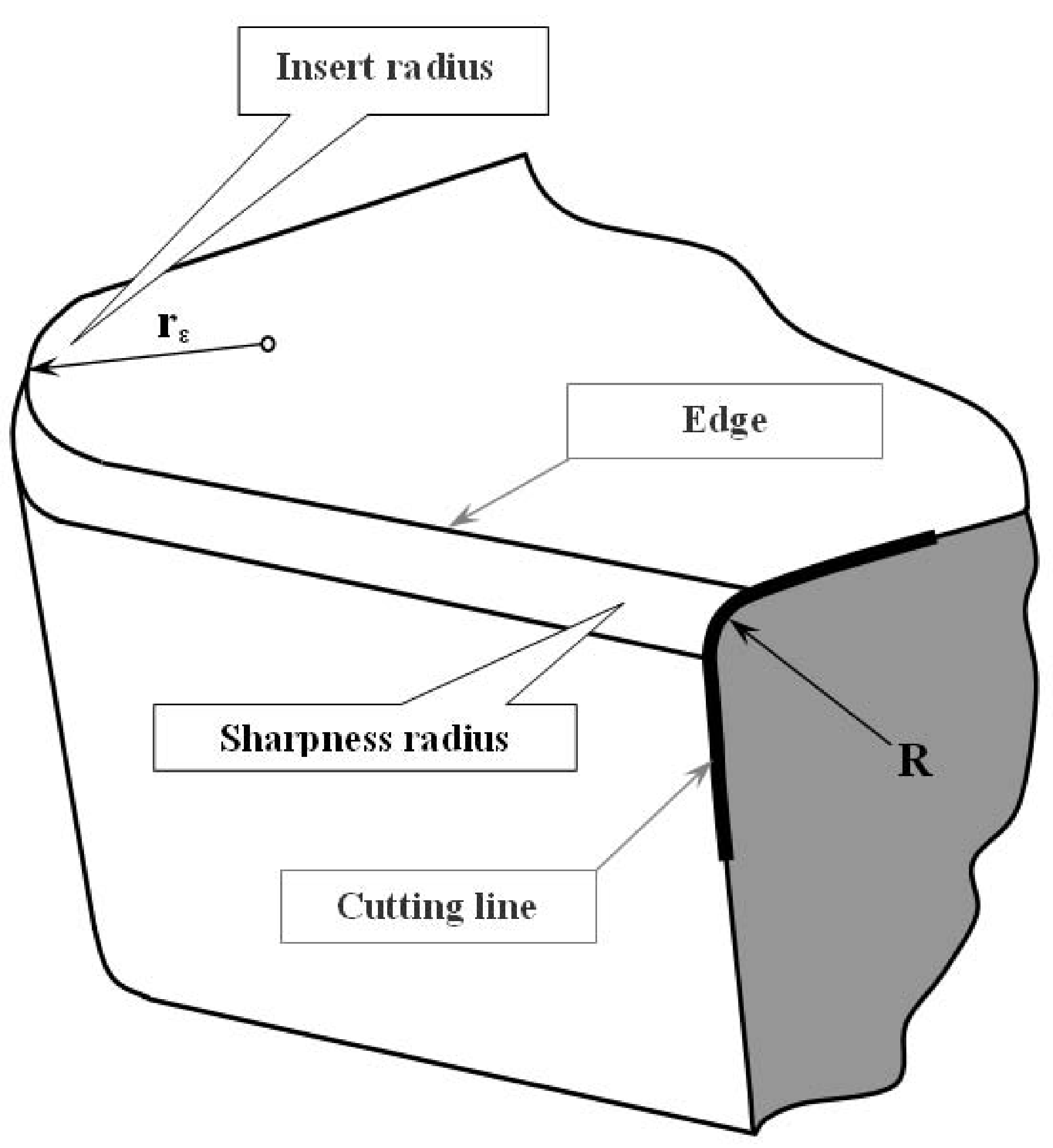}
	\caption{The tool geometrical details.}
	\label{fig5}
\end{figure}

Moreover, the geometry of the tool (Fig.~\ref{fig5}) \cite{laheurte-04} is characterized by the cutting angle $\gamma$, the clearance angle $\alpha$, the edge angle of inclination $\lambda_{s}$, the direct angle $\kappa_{r}$, the nozzle radius $r_{\epsilon}$ and the sharpness radius $R$. The tool insert is examined after each test and is changed if necessary of wear along the cutting face (Vb $\leq$ 0.2 mm ISO 3685), that may disturb the studied phenomenon. The tool characteristics used are presented in the Table~\ref{tabl-1}.

\begin{table}[htbp]
\centering
		\begin{tabular}{|c|c|c|c|c|c|}
\hline
 $\gamma$ & $\alpha$ & $\lambda_{s}$ & $\kappa_{r}$ & $r_{\epsilon}$ & R\\
\hline
$-6^{\circ}$ & $6^{\circ}$ & $-6^{\circ}$ & $91^{\circ}$ & 1,2 mm & 0,02 mm\\
\hline
\end{tabular}
\caption{The tool geometrical characteristics}
	\label{tabl-1}
\end{table}
 
\begin{figure}[htbp]
\centering
		\includegraphics[width=0.46\textwidth]{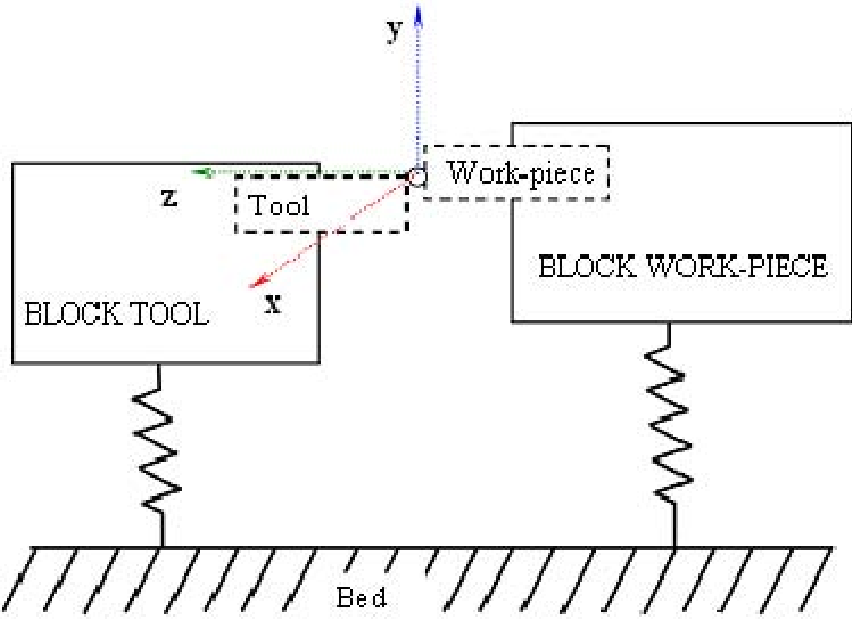}
		\caption{Model of machining system and \textbf{BW} - \textbf{BT} interaction}
	\label{fig6}
\end{figure}

Remember that the assembly of \textbf{BW} detailed in \cite{bisu-AAAA-ispas-06a} is considered as a block of interdependent elements, which is subjected to the vibrations due to the cutting process. The machining system is presented as an elastic system and the interface \textbf{BW} / \textbf{BT} is the closure of this system (Fig.~\ref{fig6}). The vibrations appearance is strongly conditioned by the elastic structure behaviour of the studied system \cite{bisu-AAAAA-ispas-07}.

Next research step is to pass to the precise analysis of the different parameters measured.

\section{Tests and analyses}
\label{sec:3}

\subsection{Experimental conditions}
\label{conditions}

Many tests are carried out under the conditions specified in the previous paragraph. They are considered for the depth of cut ap = 1~mm, 2~mm, 3~mm and 5~mm in order to compare signals and to dissociate the machine vibration influence of those generated by the cutting process itself.

\subsection{Workpiece/Tool/Chip displacements}
\label{deplacements}

\begin{figure}[htbp]
\centering
	\includegraphics[width=0.49\textwidth]{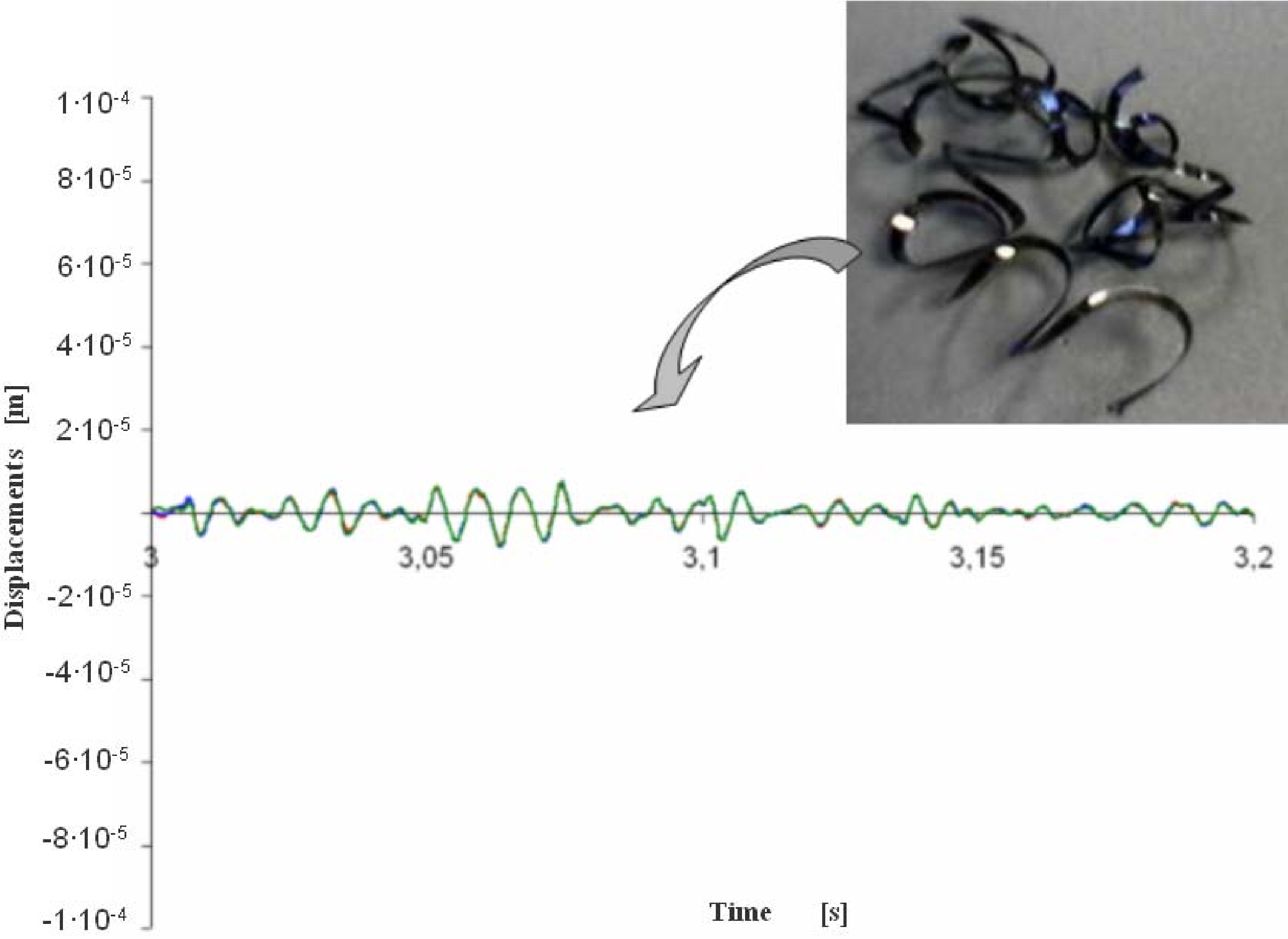}
	\caption{Displacements components signals of the tool tip for a turning operation: ap = 2~mm, f = 0.1~mm/rev, and the spindle speed N = 690~rpm}
	\label{fig7}
\end{figure}

\begin{figure}[htbp]
\centering
		\includegraphics[width=0.48\textwidth]{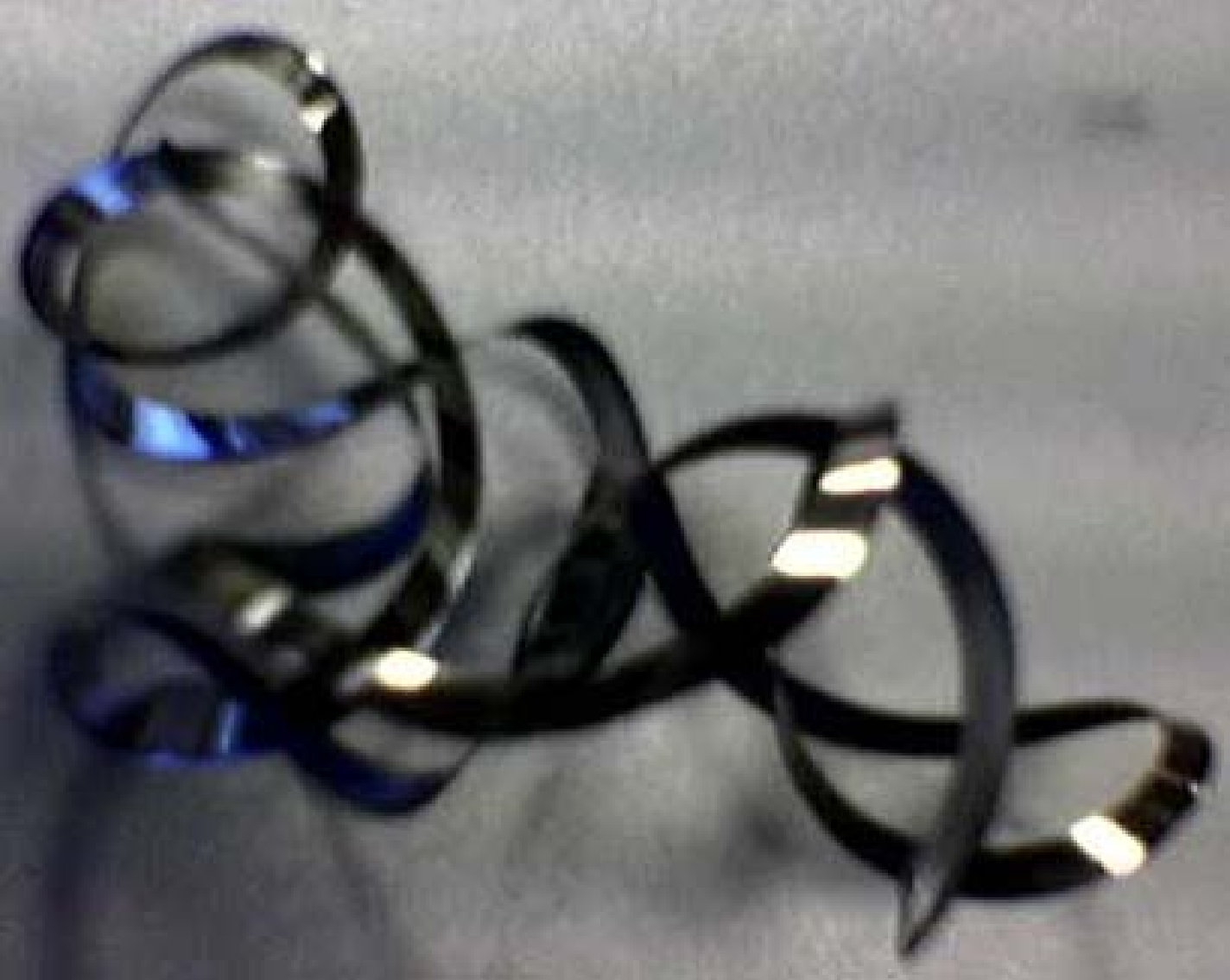}
	\caption{Chip related to ap = 3~mm, f = 0.1~mm/rev and the spindle speed N = 690~rpm}
	\label{fig8}
\end{figure}

The signals treatment of the three-direction accelerometers gives, by integration, displacements according to the three directions. An example is presented in the Fig.~\ref{fig7} related to a cutting depth of 2~mm. For these tested conditions, the system is stable; the signal has very low amplitudes, on the order of micrometer; and the chip section remains constant (Fig.~\ref{fig7} and \ref{fig8}). The part surface quality is correct, with a total roughness $(R_{t})$ of 4.3~$\mu m$.

The analysis of measurements for the depth of 3~mm cut, give similar results; the displacements amplitudes are in the same region, lower than 10 micrometers, and the chip section does not vary (Fig.~\ref{fig8}); this one has the same continuity characteristics as for ap = 2~mm (Fig.~\ref{fig7}),  but very light undulations appear on the workpiece surface.

\begin{figure}[htbp]
\centering
		\includegraphics[width=0.48\textwidth]{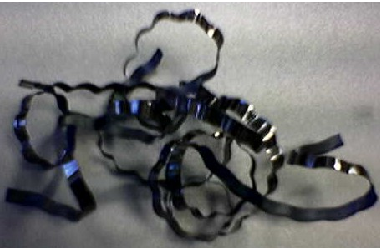}
	\caption{Chip related to ap = 4~mm, f = 0.1~mm/rev , and the spindle speed N = 690~rpm.}
	\label{fig9}
\end{figure}

Actually, the characteristics start to change distinctly from the depth of 4~mm cut. The tool point displacements amplitudes increase, and the signal sinusoidal character has high amplitude compared to the previous cases. The system starts to become unstable. The chip develops important periodic undulations and significant section variations appear (Fig.~\ref{fig9}). All chips are type 1.3 (ISO 3685). The machined surface part has weak undulations.

The increase of cutting depth until 5~mm allows to reach a mode of totally unstable cutting and the self-excited vibration appearance is clearly identified for ap = 5~mm (Fig.~\ref{fig11}). Around this working point, many tests are carried out in order to determine the self-excited vibrations extent. These tests are carried out maintaining the same ap value and the same workpiece revolutions number N but for various feed motion f, illustrated in the Table~\ref{tabl-2}.

\begin{table}[htbp]
	\centering
		\begin{tabular}{|c|c|c|c|c|c|}
\hline
ap(mm) & N (rpm) & \multicolumn{4}{|c|} {f(mm/rev)}\\
\hline
5 & 690 & 0.05 & 0.0625 & 0.075 & 0.1\\
\hline
	\end{tabular}
\caption{The cutting process parameters}
	\label{tabl-2}
\end{table}

It is important to specify that, for the four feed values used (Table~\ref{tabl-2}), the cutting process was in unstable mode, more or less marked, with chip section variations and very different surface undulations on the workpiece for each imposed feed value. In order to reach the best possible precision, we then study the extreme case obtained with the maximum values which involve the most important dynamic effects: cutting depth ap = 5~mm, and feed motion value f = 0,1~mm/rev, but maintaining a constant cutting speed.

\begin{figure}[htbp]
\centering
		\includegraphics[width=0.48\textwidth]{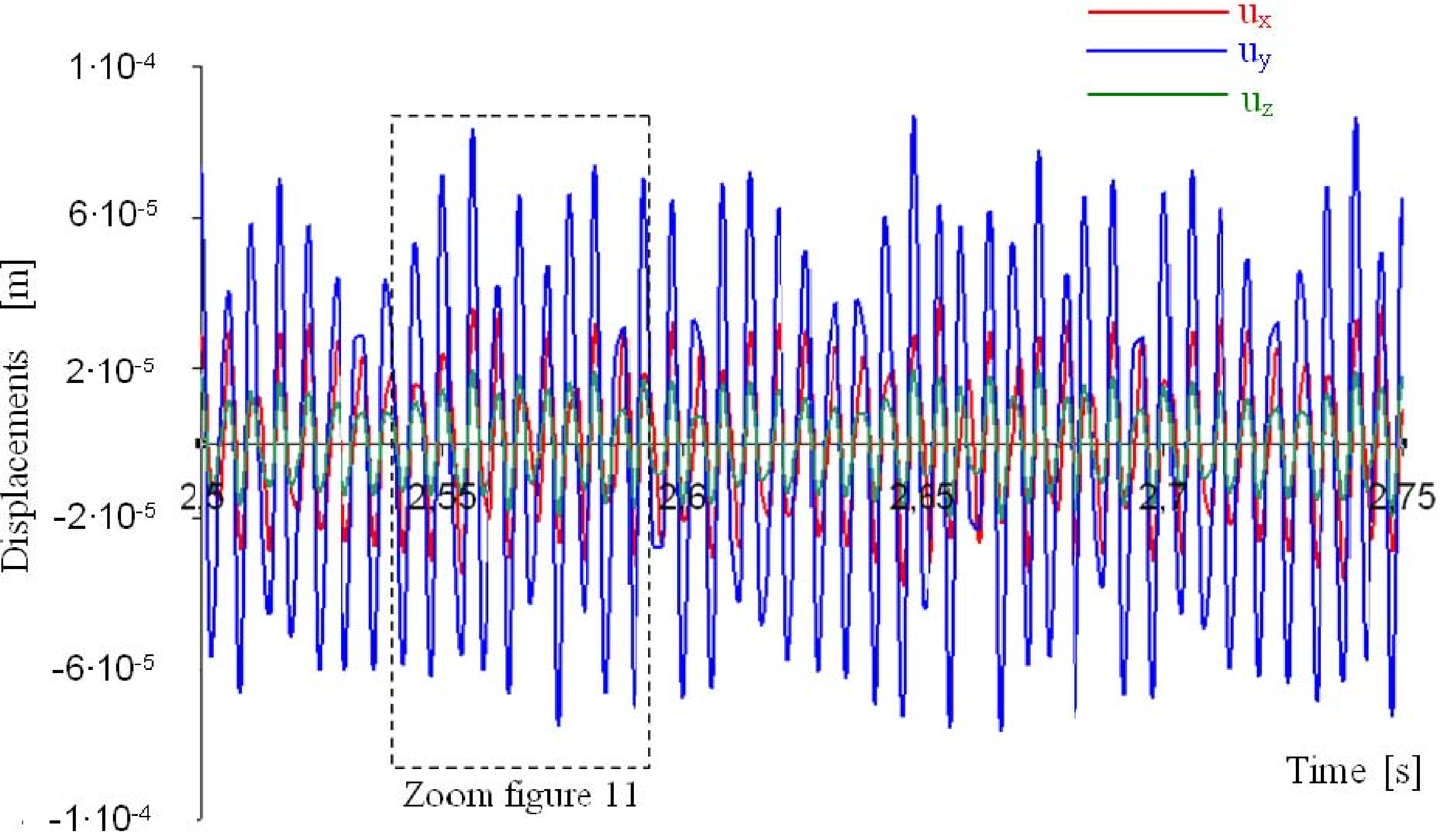}
	\caption{The tool tip displacements on the three directions x, y, and z related to the machine tool and considering ap = 5~mm, f = 0,1~mm/rev, N = 690~rpm}
	\label{fig10}
\end{figure}

\begin{figure}[htbp]
\centering
		\includegraphics[width=0.49\textwidth]{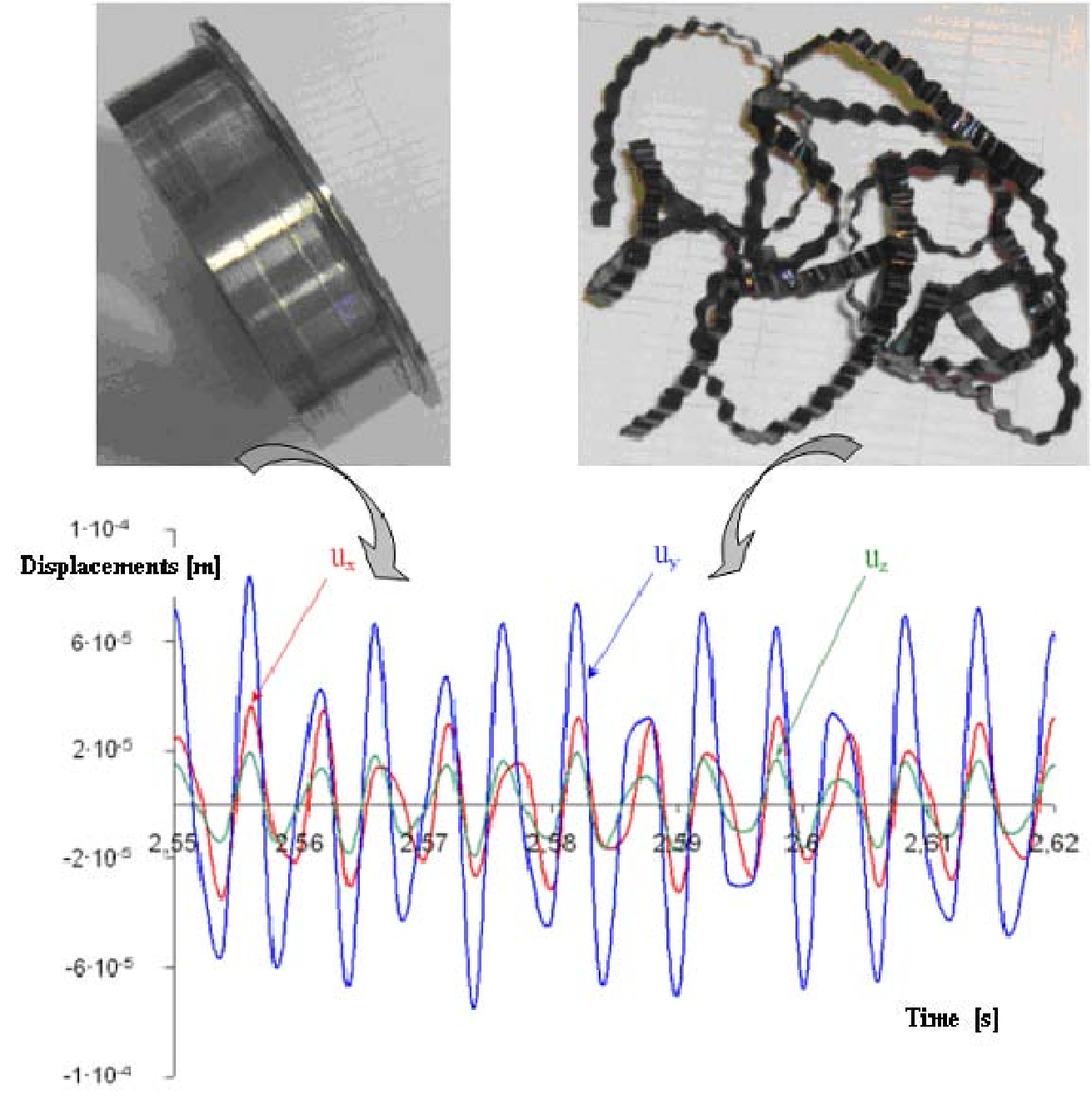}
	\caption{Signals zoom related to the tool tip displacements components according with cutting parameters: ap = 5~mm, f = 0,1~mm/rev., N = 690~rpm}
	\label{fig11}
\end{figure}

In Fig.~\ref{fig10} are presented the three signals related to the tool tip displacements components according to the three directions of workspace.

A signals zoom related to the tool tip displacement components is presented in the Fig.~\ref{fig11}, as well as the results of cutting process vibrations on the workpiece and chip.

In this situation, the chip presents important tightened periodic undulations. The section variations seem to be in the case of a discontinuous cutting. The workpiece surface quality becomes poor and the undulations left on the workpiece after the  tool passage are very important (Fig.~\ref{fig11}). The chip characterizations and the workpiece are detailed in \cite{bisu-07}; in particular, the undulations and the chip section variations are quantified there.

\subsection{Displacements frequency analysis related to the tool tip}
\label{analyse}

The tool tip displacements components frequency analysis considering the three directions related to the machine tool is carried out starting from a Fast time Fourier Transformation (FFT). An FFT spectrum example of the three acceleration components delivered by the three-directional accelerometer placed on the tool is given in the Fig.~\ref{fig11}. The frequencies above 1~000~Hz are eliminated knowing that, according to the literature, the self-excited vibration frequency domain is situated between 120~Hz and 200~Hz \cite{ispas-AA-anghel-99}, \cite{kudinov-70}, \cite{marot-80}, \cite{martin-73}, \cite{moraru-A-rusu-79}, \cite{tansel-A-keramidas-92}.

The FFT analysis of several tests shows that the acceleration frequencies of vibrations during the cutting process are about 150~Hz to 200~Hz. They are slightly lower than the  machine natural frequencies (Fig.~\ref{fig3}). However, we know that the self-excited vibrations appear near the natural and the machining system are not depending on the number of workpiece revolutions \cite{ispas-AAA-boboc-98}, \cite{moraru-A-rusu-79}.

\begin{figure}[htbp]
\centering
		\includegraphics[width=0.48\textwidth]{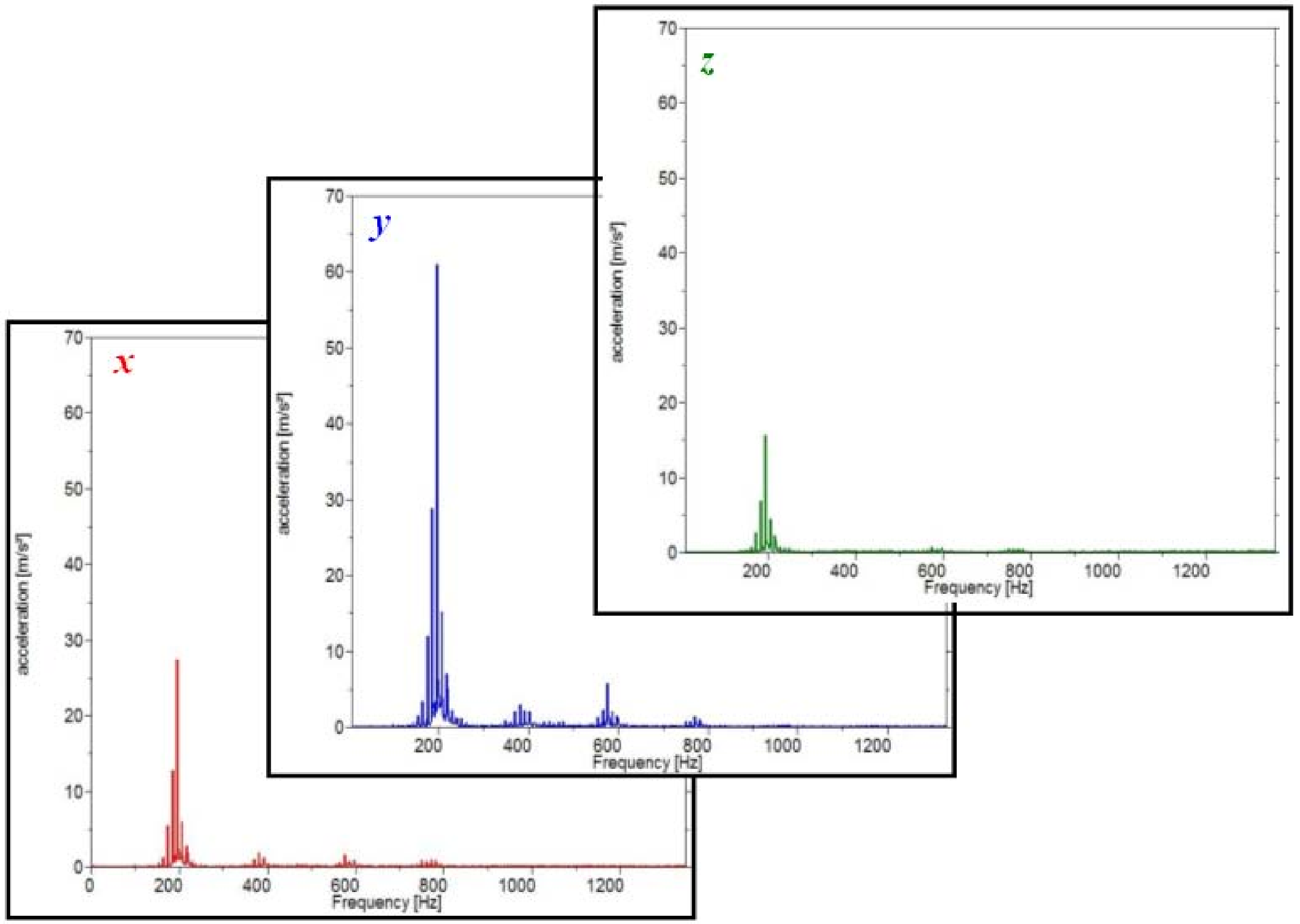}
	\caption{FFT spectrum corresponding of the accelerations acquires on the three directions; cutting parameters selected ap = 5~mm, f = 0.1~mm/rev. and N = 690~rpm}
	\label{fig12}
\end{figure}

On the example of vibrations measurements during the machining, presented in the Fig.~\ref{fig12}, it should be noted that the cutting vibrations fundamental frequencies are around 190~Hz for the three axes, with a dominating amplitude on the y axis (cutting axis, according to Fig.~\ref{fig2}). These frequencies are lower than those obtained (\cite{bisu-AA-k'nevez-07}) during the machining system dynamic characterization with a no-cutting process, which are located at approximately 550~Hz (Fig.~\ref{fig3}) \cite{bisu-AAAAA-ispas-07a}. The recorded frequencies, therefore, correspond well to the self-excited vibrations related to the cutting process. The frequency peak around 190~Hz is found for the other tests, and it is constant according to the feed motion in turning.

Another very important parameter for the self-excited vibrations is their amplitude. It depends on the energy quantity introduced into the system. Therefore, the amplitude is influenced on the one hand by the system elastic parameters, and, on the one other hand, by the cutting parameters: cutting speed, feed motion and depth of cut. The parameter that presents the most important influence is the depth of cut. The vibrations amplitude increases in a quasi linear way according to the increase depth of cut until it exceeds the elastic system stability limit in agreement with literature data published by Kudinov \cite{kudinov-70}.

\section{Tool point displacements plane}
\label{sec:4}

\subsection{Acceleration analysis of vibration data}
\label{donneesaccelero}

\begin{figure}[htbp]
	\centering
		\includegraphics[width=0.48\textwidth]{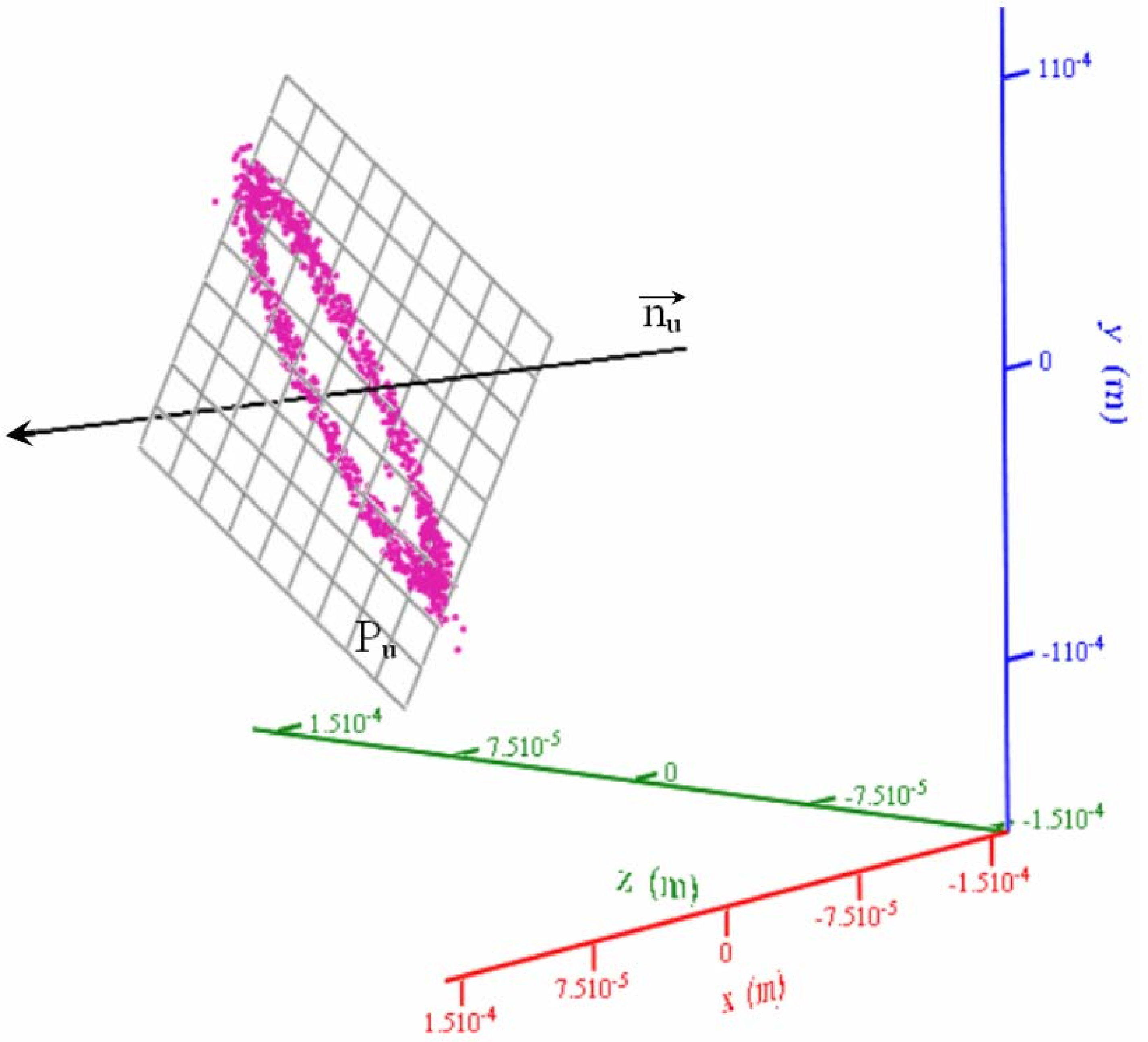}
	\caption{The tool tip displacements plane $P_{u}$}
	\label{fig13}
\end{figure}

After integration, the acceleration data analysis allows to establish the existence of an attached displacements plane of the tool tip, which, describes an ellipse (Fig.~\ref{fig13}). Certain parameters are a slightly increasing depending of the feed motion, as we will demonstrate further.

The determination of this plane is presented in appendix (Sect:~\ref{sec:DEterminationDuPlanDesDEplacements}). The existence of this plane is confirmed by the displacements signals observation (Fig.~\ref{fig11}). It is henceforward a question of locating this plane in three dimension space. 

\subsection{The tool tip displacements plane localization}
\label{localisation}

\begin{figure}[htbp]
	\centering
		\includegraphics[width=0.48\textwidth]{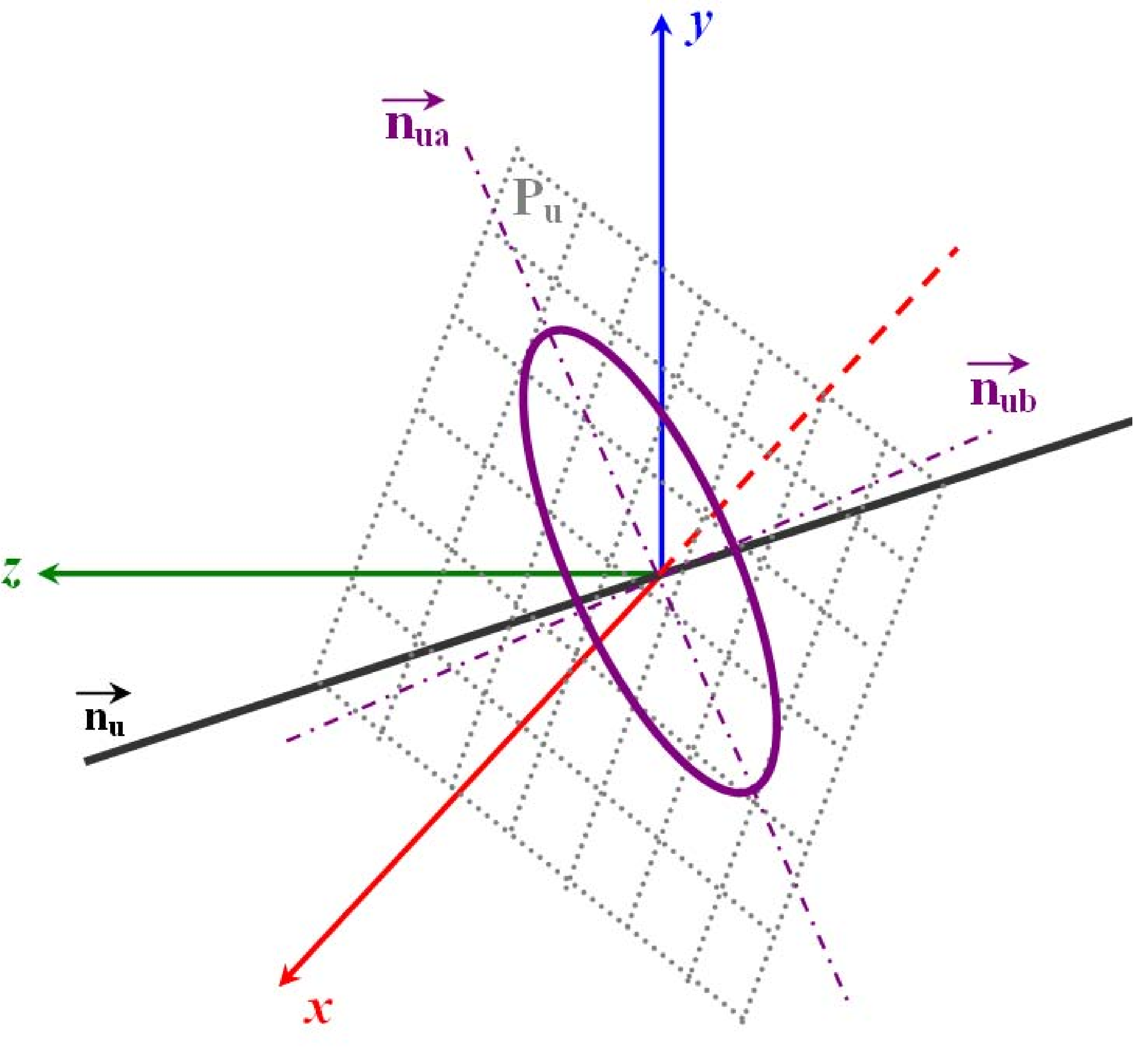}
	\caption{The tool tip displacements plane $P_{u}$: projection of the normal vector $\vec{n_{u}}$ on the axis related to the machine tool}
	\label{fig14}
\end{figure}

The tool tip displacements plane $P_{u}$ is characterized by its normal vector noted $\vec{n_{u}}$ (Fig.~\ref{fig14}).

The characterization of this plane by its normal is carried out for each test depending on the feed motion in turning (Table~\ref{tabl-3}).

\begin{table}[h!tbp]
	\centering
		\begin{tabular}{|c|c|c|c|c|}
\hline
f(mm/rev)& 0.05 & 0.0625 & 0.075 & 0.1 \\
\hline
Normal & \multicolumn{4}{|c|}{$\vec{n_{u}}$}\\
\hline
x direction & -0.071 & -0.071 & -0.058 &-0.056\\
\hline 
y direction & -0.186 & -0.186 & -0.206 & -0.216\\
\hline
z direction & 0.98 & 0.98 & 0.977 &  0.975\\
\hline
		\end{tabular}      
\caption{The normal $\vec{n_{u}}$ of the plane $P_{u}$ attached of the tool tip displacements on three directions of cutting, depending on feed motion}
	\label{tabl-3}
\end{table}

We note that the normal component according to the feed motion direction almost does not evolve according to this one. On the other hand, a very light decrease (respectively growth) according to the radial direction (respectively, the axis of cutting) is observed in dependence with the feed motion values. These variations are, however, so weak that they do not affect the standard of the plane direct normal which thus can be considered constant according to the feed motion. The existence and the determination of this plane provide simplifying important informations about the configuration to be adopted in order to write a semianalytical model that is under development \cite{bisu-07}, \cite{cahuc-A-laheurte-07}. In particular, we will further see that this plane allows to bring back the cutting three-dimensional problem, with three directions displacements, using a simpler plane model that is inclined compared to axes (x, y, z) related to the machine tool.

\section{Displacements analysis of the tool point}
\label{sec:5}

\subsection{Stable and unstable process comparison}
\label{ComparSatbl} 

\begin{figure}[htbp]
	\centering
		\includegraphics[width=0.48\textwidth]{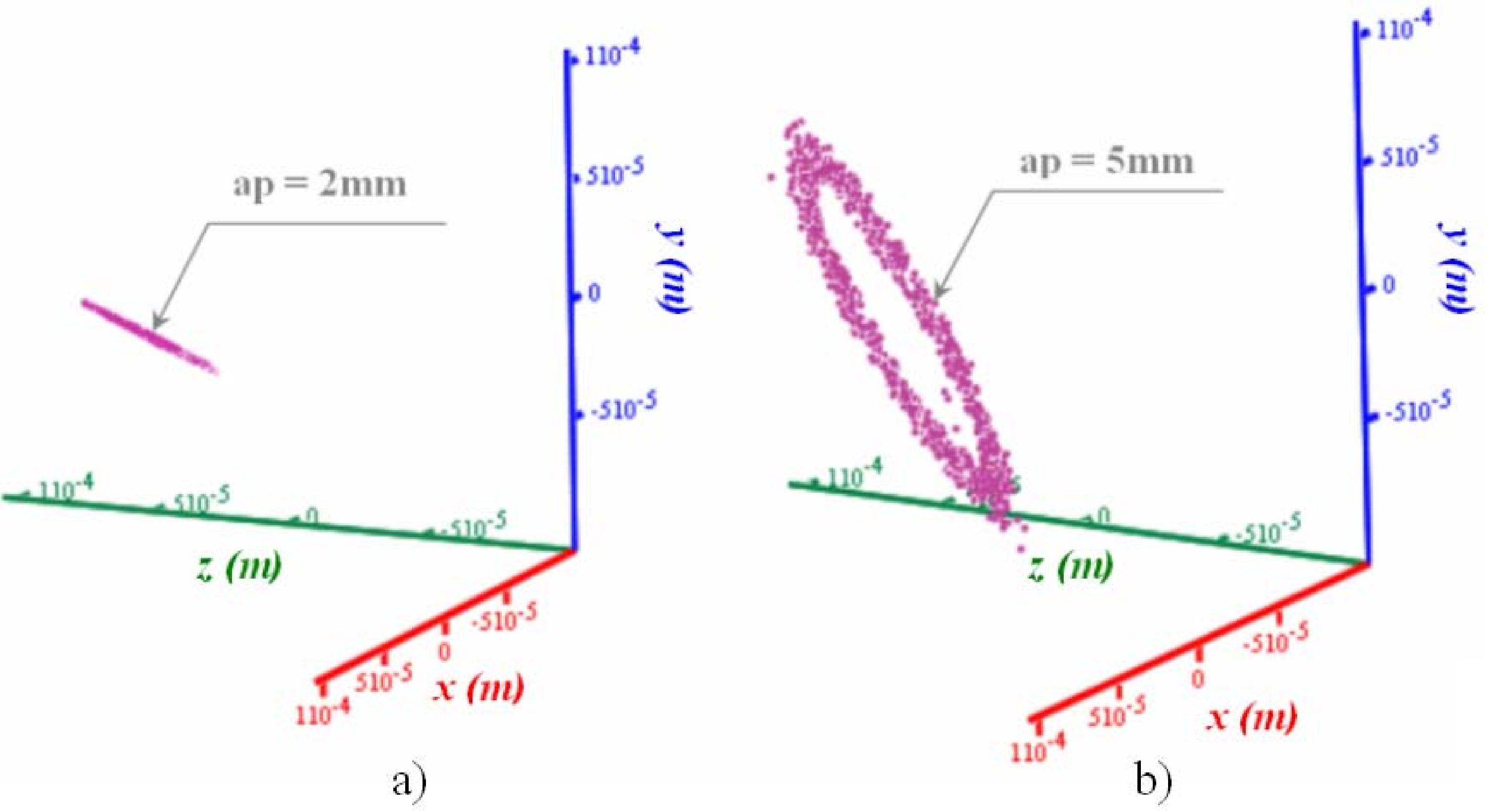}
	\caption{Comparison between the stable (a) and unstable cutting process (b)}
	\label{fig15}
\end{figure}

In the tool tip displacements plane $P_{u}$, it is possible to establish that the tool tip point describes a very small ellipse in the stable case cutting process (without vibrations); this ellipse could be assimilate with a small segment of straight lines (ap =~2~mm), according to the Fig.~\ref{fig15}-a. On the other hand, in unstable mode (with vibrations), the ellipse (according~\ref{sec:ApproximationDeLEllipse}) of tool tip displacements (ap = 5~mm) is really more visible (Fig.~\ref{fig15}-b).

\subsection{Phase control between the tool tip displacements components}
\label{dephas}

\begin{figure}[htbp]
	\centering
		\includegraphics[width=0.48\textwidth]{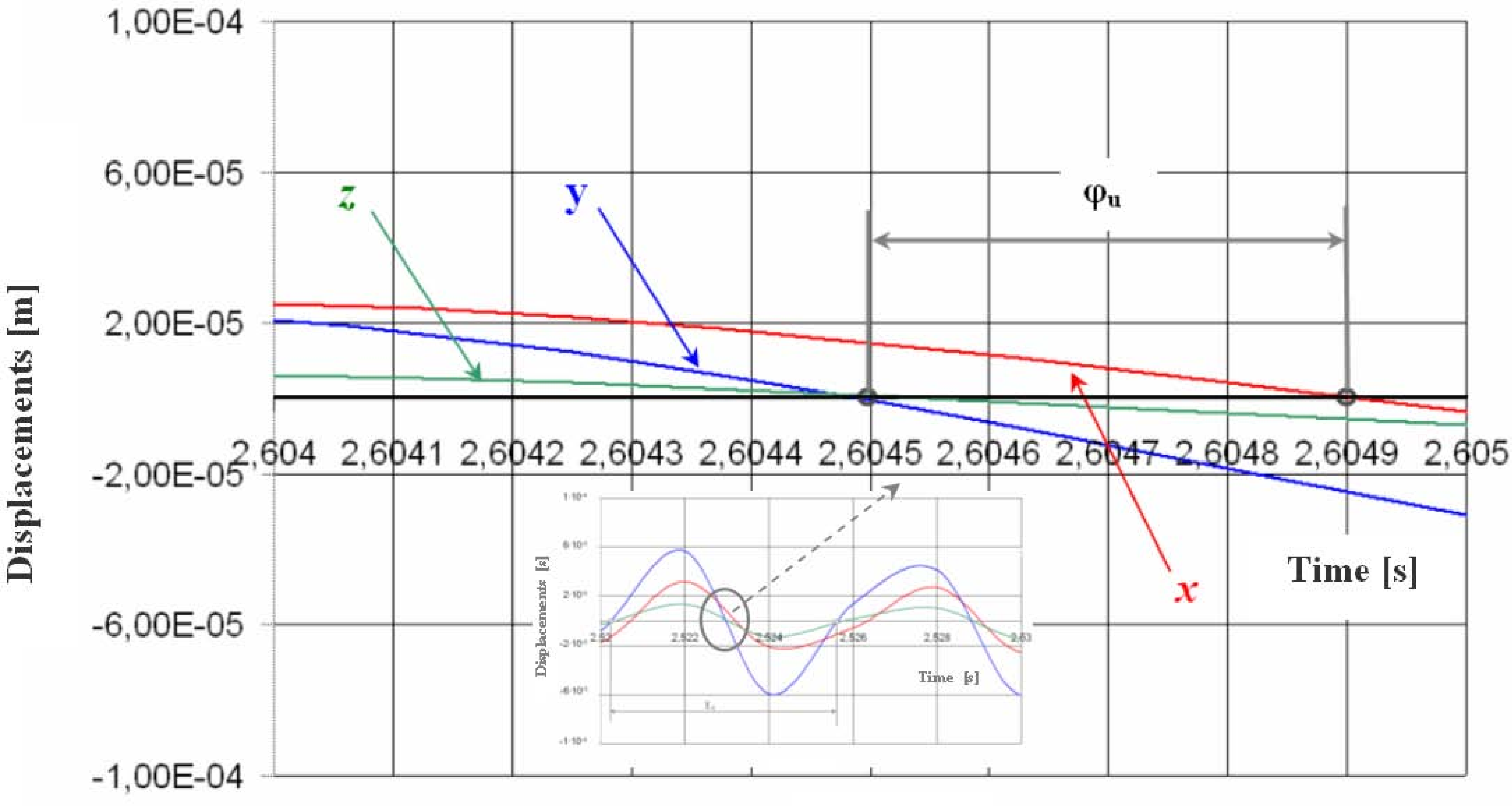}
	\caption{Example of phase difference between $x$ radial tool tip displacements and other displacement components (following cutting direction and feed motion direction).}
	\label{fig16}
\end{figure}

A phase difference is occurring between the tool tip displacements following the $x$ radial direction and the $y$ cutting direction or the $z$ feed motion direction (these last two being almost in phase). This phase difference is evaluated starting from the comparison of two signals that have different phases. More precisely, phase difference $\varphi_{u}$ is evaluated starting directly to the temporal signals and using the relation: 

\begin{equation}
\varphi_{u}=\pm2\pi\frac{\Delta t}{T},
\end{equation}

where $T$ is the time corresponding to a phase difference of $360^{\circ}$ and ${\Delta t}$ to temporal phase difference between the two compared components signals. Fig.~\ref{fig16} presents an example of phase difference $\varphi_{u}$ between radial displacement $x$ and the two displacements components following $y$ and $z$; they are almost in phase. A phase difference of $28^{\circ}$ is thus obtained for the studied case, ap =~5~mm and f =~0.1~mm/rev. 

During the experimental tests according to different feed motion values, the variation of this phase difference is very weak. For the value f= 0,05~mm/rev, the phase difference remains equal to the phase difference obtained for the feed value f =~0,1~mm/rev.

\subsection{Tool tip displacements ellipse approximation}
\label{elldepl}

\begin{figure}[htbp]
	\centering
		\includegraphics[width=0.42\textwidth]{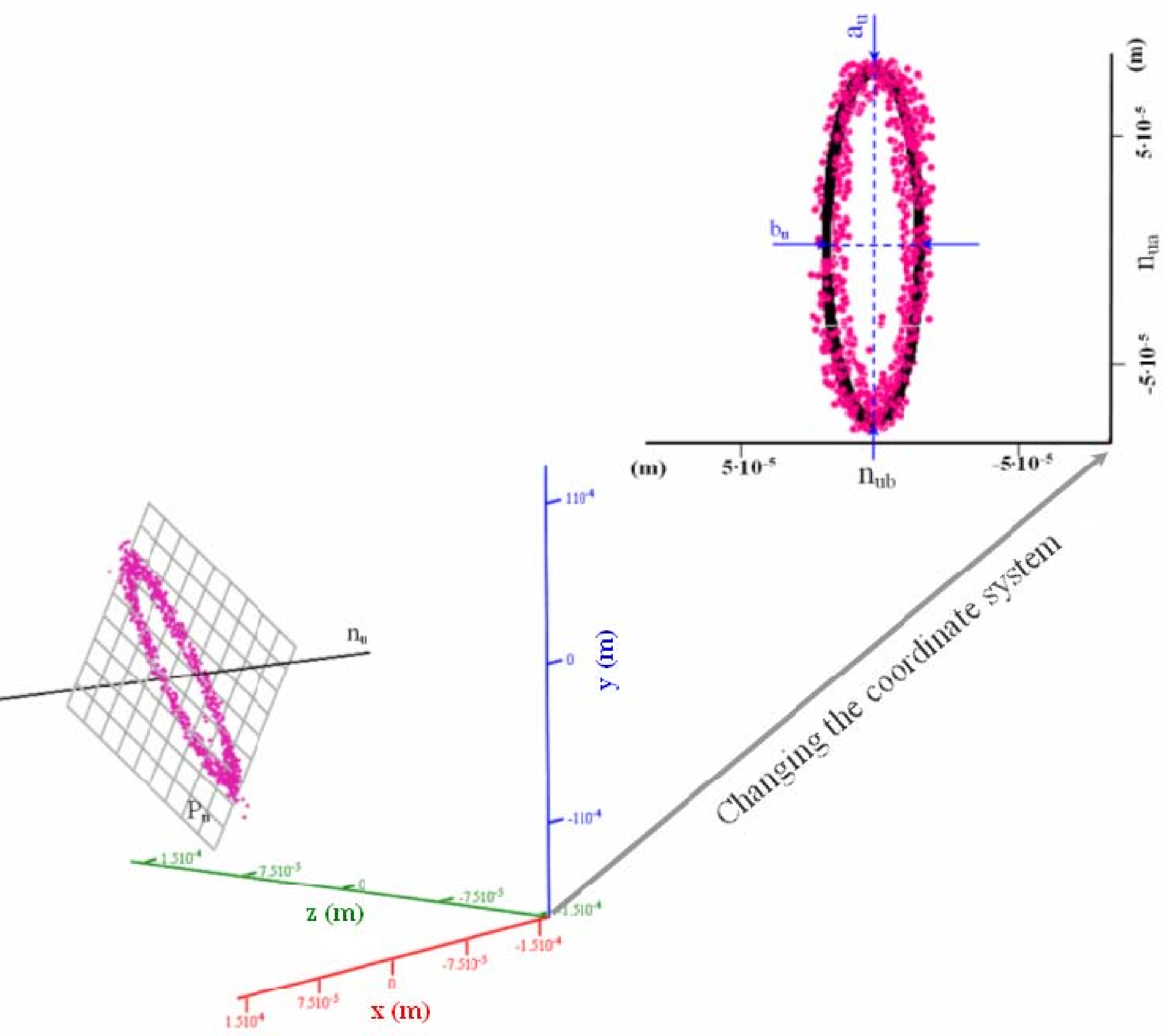}
	\caption{Coordinate system changing to obtain the ellipse attached to the tool point displacements}
	\label{fig17}
\end{figure}

The ellipse is characterized in detail in the appendix (cf. Sect:~\ref{sec:ApproximationDeLEllipse}) on the plane $P_{u}$, with the normal direction $\vec{n_{u}}$, by its large axis $a_{u}$ and its small axis $b_{u}$ determined using the method of least squares, in the reference system $\left(\vec{n_{ua}},\vec{n_{ub}}\right)$ unit vector related to the ellipse axes (Fig. ~\ref{fig17}).
 
The values of $a_{u}$, $b_{u}$ and their ratio  $a_{u}/b_{u}$ are presented in the Table~\ref{tabl-4}. The value analysis of the large and small ellipse axis shows that these two sizes are increasing with the feed value. On the other hand, a weak growth (6\%) of the ratio $a_{u}/b_{u}$ is obtained when the feed value decreases. The ellipse geometrical proportions are thus quasiconstant according to the feed weak evolution chosen here.

\begin{table}[htbp]
	\centering
		\begin{tabular}{|c|c|c|c|}
\hline
f(mm/rev)&$a_{u}$ (m)&$b_{u}$ (m)&$a_{u}$/$b_{u}$\\
\hline
0.1 & $7.71.10^{-5}$ & $1.72.10^{-5}$ & 4.48\\
\hline
0.075 & $5.6.10^{-5}$ & $1.23.10^{-5}$ & 4.59\\
\hline
0.0625 & $3.87.10^{-5}$ & $0.83.10^{-5}$ & 4.66 \\
\hline
0.05 & $2.83.10^{-5}$ & $0.51.10^{-5}$ & 4.76\\
\hline
\end{tabular}
\caption{Values depending on the feed for large and small ellipse axes that correspond to tool tip displacements for ap = 5~mm, f = 0.1~mm/rev}
	\label{tabl-4}
\end{table}

\section{Discussion on static and dynamic aspects}
\label{sec:6}

\subsection{Stiffness / displacements correlation}
\label{correlDepla}

Taking into account the machining system static study realized in \cite{bisu-AAAA-ispas-06a}, it is possible to check that the stiffness principal directions angles are comparable with the ellipses angles attached to the tool tip displacements (Table~\ref{tabl-5}).

\begin{table}[h!tbp]
	\centering
\begin{tabular}{|c|c|c|c|c|}
\hline
\multicolumn{2}{|c|}{static case}& \multicolumn{3}{|c|}{dynamic case} \\
\hline
$\alpha_{K(yz)}$ & $\alpha_{K(xy)}$ & $\theta_{e(yz)}$ & $\theta_{e(xy)}$ & f (mm/rev)\\
\hline
 &  \multirow{4}* & $77^{\circ}$ & $69^{\circ}$& $0.1$ \\
$76^{\circ}$ & $65^{\circ}$ & $77^{\circ}$ & $69^{\circ}$& $0.075$ \\
& & $79^{\circ}$ & $67^{\circ}$ & $0.0625$ \\
& & $79^{\circ}$ & $65^{\circ}$ & $0.05$ \\
\hline

		\end{tabular}
\caption{Comparison values between the principal directions angles of the tool tip displacements and the stiffness principal directions of the machining system}

	\label{tabl-5}
\end{table}

Between the vector $\vec{n_{f}}$ (characterizing the stiffness principal directions) and the normal vector $\vec{n_{u}}$ situated in the attached tool tip displacements plane, another correlation exists. Indeed, calculation shows that the angle between these two vectors does not exceed $2.5^{\circ}$, in our case (Table~\ref{tabl-6}). It is advisable, moreover, to note that the difference between the two vectors is slightly increasing depending on the feed value.

\begin{table}[htbp]
	\centering
	\begin{tabular}{|c|c|}
\hline
f(mm/rev)&$\arccos\left(\frac{\vec{n_{f}}.\vec{n_{u}}}{\left\|\vec{n_{f}}\right\|.\left\|\vec{n_{u}}\right|\|}\right)$\\
\hline
0.1 & $2.5^{\circ}$\\
\hline
0.075 & $1.9^{\circ}$\\
\hline
0.0625 & $1^{\circ}$ \\
\hline
0.05 & $0.8^{\circ}$\\
\hline
	\end{tabular}
	\caption{Correlation between $\vec{n_{f}}$  - stiffness principal directions and $\vec{n_{u}}$ - normal direction of the plane $P_{u}$ attached of the tool tip displacements}
	\label{tabl-6}
\end{table}
	
\subsection{Correlation between tool displacements / stiffness center}
\label{correlRaid}

By comparison with the static analysis results developed in \cite{bisu-07}, it is highlighted a coupling between the system elastic characteristics \textbf{BW} and the vibrations generated by cutting process. As we could expect it, the self-excited vibrations appearance is strongly influenced by the system stiffness values, their ratio and their direction.

\begin{figure}[htbp]
	\centering
		\includegraphics[width=0.48\textwidth]{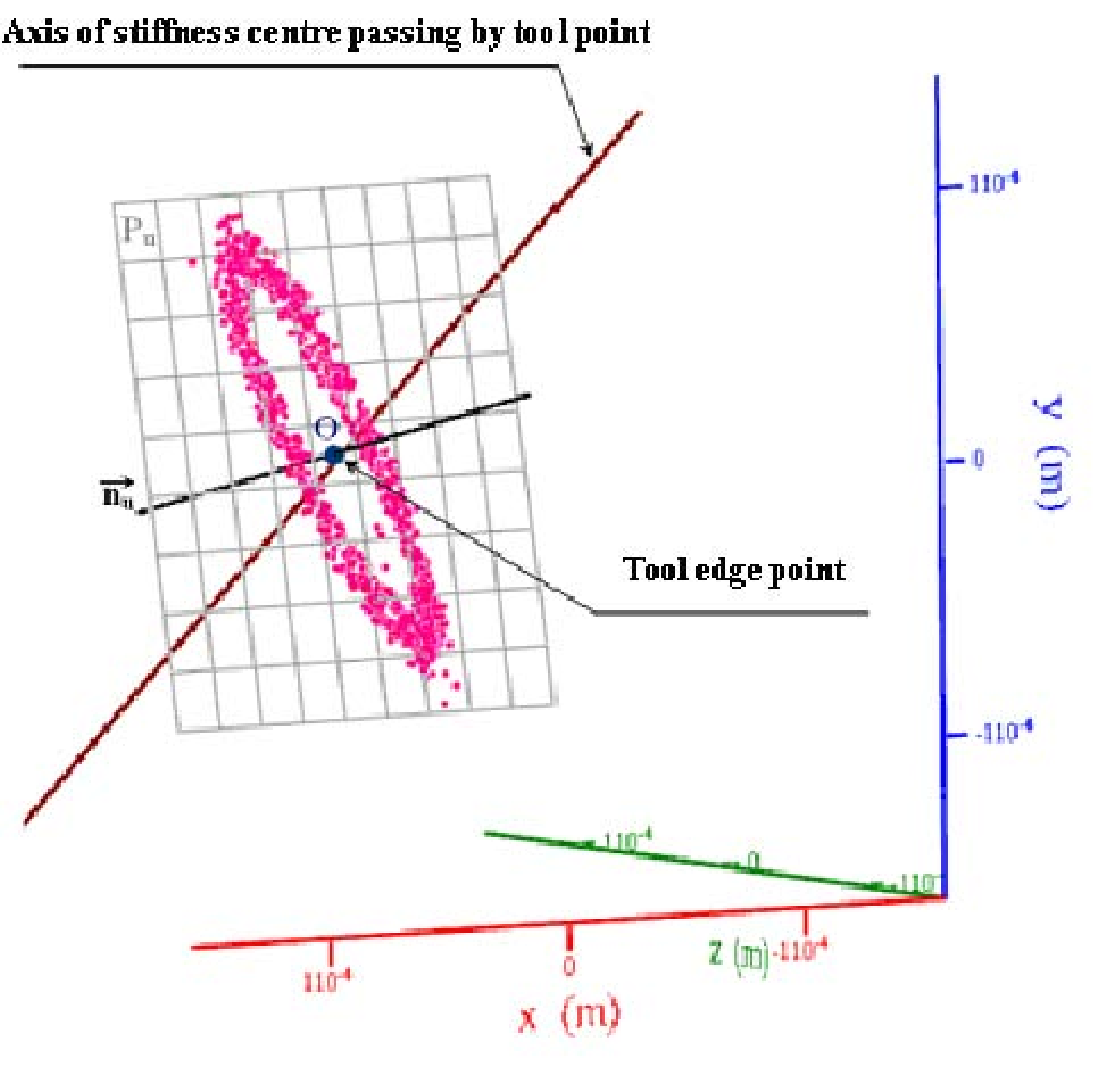}
	\caption{Correlation between stiffness center and the ellipse attached of the tool tip displacements }
	\label{fig18}
\end{figure}

The stiffness center (CR) is given in \cite{bisu-AAAA-ispas-06a} (and a summary of the stiffness center determination is given in appendix \ref{sec:stiffneesCenter}). The axis, which passes by the stiffness center and the tool tip, represents the maximum stiffness direction. This maximum stiffness axis is drawn in Fig.~\ref{fig18}. It is orthogonal with the normal of the plane $P_{u}$, and it is also perpendicular to the large ellipse axis. This explains well why the greatest displacement takes place, logically, on the axis where the stiffness is minimal. This correlation validates the experimental protocol established \cite{bisu-AAAA-ispas-06a} for this case of turning.

To continue, the $P_{u}$ displacements plane determination and the elliptic characterization of this plane motion provide significant information on the configuration to be adopted writing a simplified three dimensional model. It is clear at the end of this research that the system must be considered according to this plane axes and not according to those of the machine tool.

\subsection{Correlation between stiffness center and central axis of the dynamic process in turning}
\label{correlDynam}

\begin{figure}[htbp]
	\centering
		\includegraphics[width=0.40\textwidth]{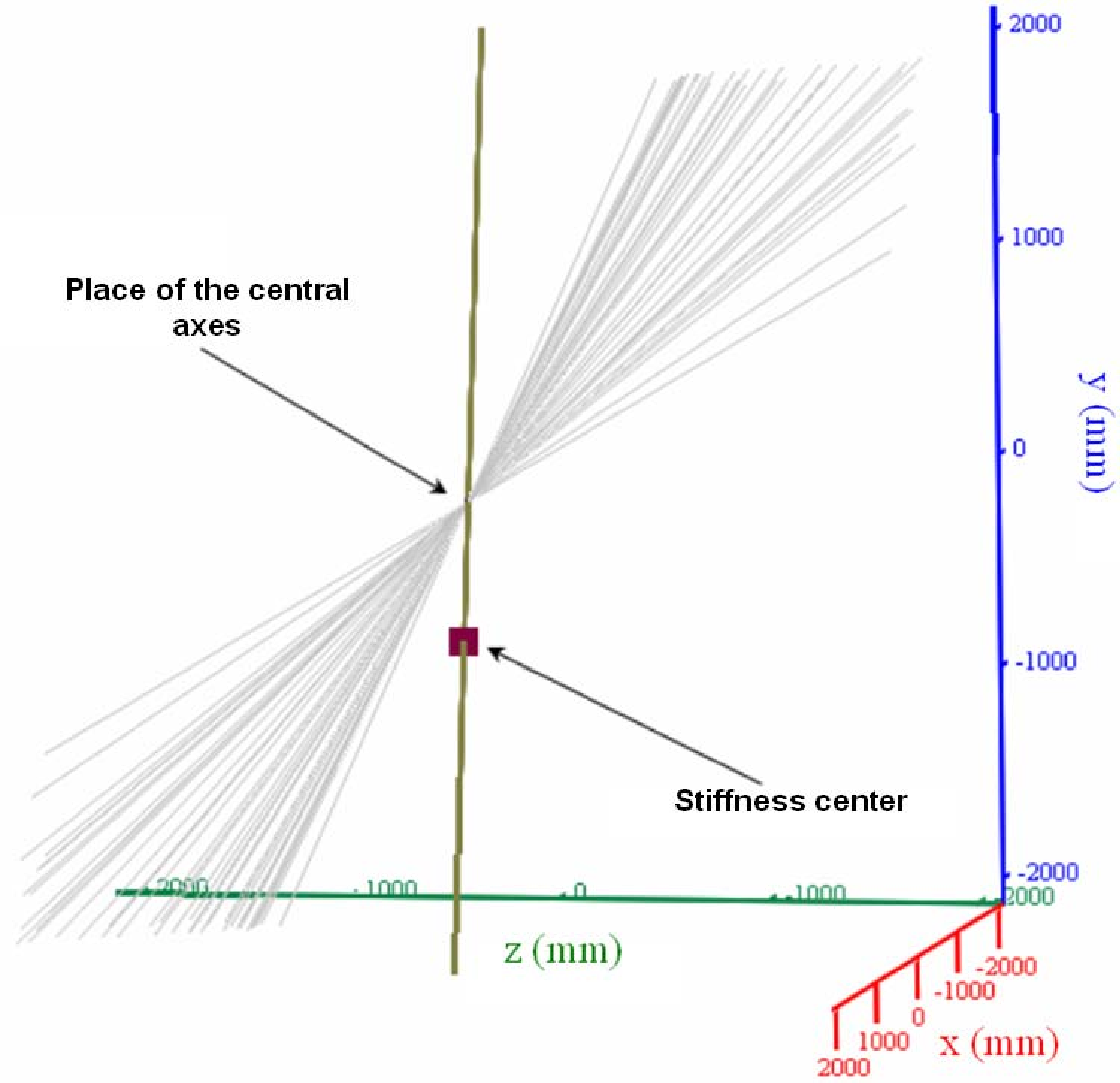}
	\caption{Central axis / stiffness center}
	\label{fig19}
\end{figure}

As example, for a bearing test on 68 revolutions (ap = 5~mm, f = 0.1~mm/rev. and N= 690 rpm) we calculated the beam of the central axis of the tip tool point small displacements torsor, their intersection point and the stiffness center corresponding to this configuration. All these results are presented on Fig.~\ref{fig19}. The line that binds the stiffness center (CR) and the point of central axes intersection passes by the tool tip point.

\section{Conclusion}
\label{conclusion}

Compared to other authors who use only the tool stiffness \cite{calderon-98}, \cite{liu-A-liang-07}, \cite{moufki-AA-rausch-00}, this study takes into account, moreover, stiffness resulting from the system fixture/tool-holder (\textbf{BT}). The important information on the static stiffness, obtained in \cite{bisu-AAAA-ispas-06a}, certainly makes it possible to determine the privileged displacements directions.

The stiffness center and the rotation center were obtained experimentally \cite{bisu-AAAA-ispas-06a}. The direction of minimum displacement was also defined starting from the experimental model. The confrontation of these results: stiffness center, principal stiffness directions, central axis and the tool tip displacements place show the coherence of the presented new approach and consolidates the experimental procedures implemented.

From the present work, the following conclusion can be drawn:
 \begin{itemize}
\item[-] The acceleration data analysis allows to establish the existence of an attached plane of the tool tip displacements tilted compared to axes related to the machine tool. 
\item[-] In this plane, the tool tip displacements describes an ellipse.
\item[-] We have established that the tool tip point describes a very small ellipse (assimilated with a straight-line small segment) in the case of the stable cutting process (without vibrations); in unstable mode (with vibrations) the ellipse of displacements is really more visible.
\item[-] A phase difference is occurring between the tool tip displacements following the radial and cutting direction or the feed motion direction (these two last being almost the same phase).
\item[-] The long and small ellipse axis values show that these two sizes are increasing with the feed value.
\item[-] A weak growth (6\%) of the ratio long on small axes is obtained when the feed value decreases. The ellipse geometrical proportions are, thus, quasiconstant according to the feed weak evolution chosen here.
\item[-] The axis which passes by the stiffness center and the tool tip represents the maximum stiffness direction.
\item[-] The maximum (resp. minimum) stiffness tool axis is perpendicular to the large (resp. small) ellipse displacements axis.
\item[-] The self-excited vibrations appearance is strongly influenced by the system stiffness values, their ratio and their direction.
\end{itemize}

It is also important to note for this research the plane determination attached to the tool tip displacements which is carried out according to an ellipse the surface of which is an increasing function depending on the feed motion value; this aspect is in perfect coherence with the power level injected in the system during machining.

\section{Appendix}
\label{sec:Annexes}

\subsection{Reminder of stiffness center determination}
\label{sec:stiffneesCenter}

It is well known that any machine tool is characterized by principal directions of deformation. These principal directions of displacement are a function of the machine structure, its geometrical configurations, and cutting parameters used. We can observe either a very stiff behavior, or very rubber band according to the direction. Here, the tool is regarded as forming integral part of block tool \textbf{BT}. The aim is to determine the stiffness center \textbf{CR}$_{BT}$ of elastic system \textbf{BT} \cite{marinescu-A-boboc-02}. This stiffness center corresponds to the block \textbf{BT} rotation center compared to the lathe bed. Obtaining \textbf{CR}$_{BT}$ consists in finding the intersection points of different perpendicular to displacements. 

\begin{figure}[htbp]
	\centering
		\includegraphics[width=0.50\textwidth]{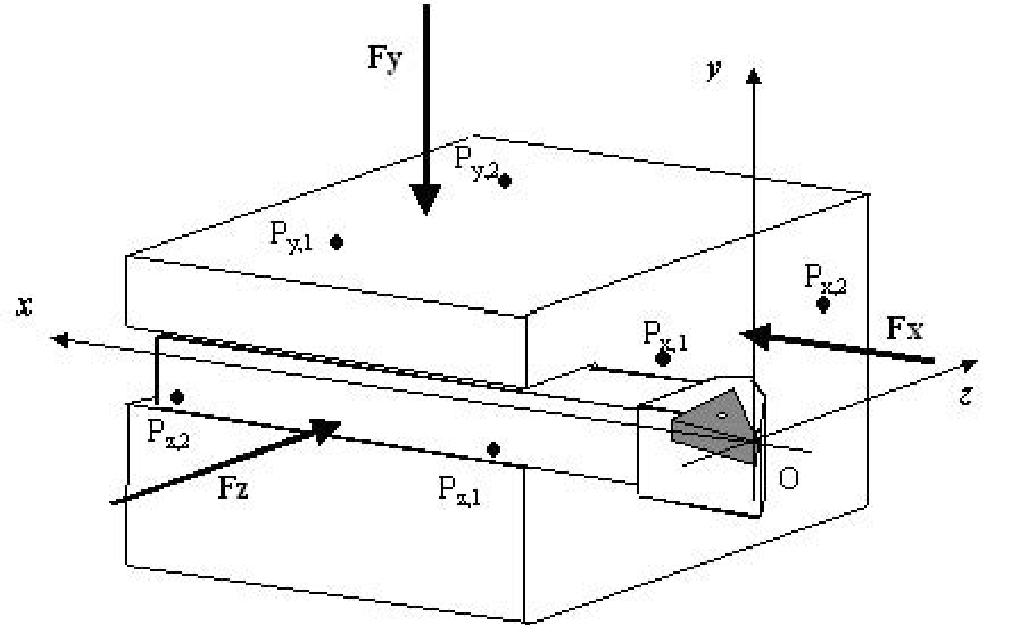}
	\caption{Experimental procedure to obtain stiffness center CR$_{BT}$}
	\label{fig20}
\end{figure}

The stiffness center is obtaining based on the tool point O that is origin of the coordinate system (x, y, z). The procedure to obtain the stiffness center is based on the virtual work principles and detailed in the figure \ref{fig20}; the imposed charge follows all three directions, Fi being known; (i = x, y, z). Measurement of the displacement vector (linear and rotation displacement) under the charge di,j is obtained in two points Pi,j in each direction (i = x, y, and z) and (j = 1 and 2). The direction (Dij) is released being straight line containing di,j vector that includes the point Pi,j .

Using the method of least squares we can minimize the distance dx between the lines. Each of points Mi on the lines (Dij) are separated by a minimum distance $\mu_{i}$ = 1,8~mm and maximum deviation $\delta_{i}$ = 2$^{\circ}$. 

In the next step, three mean plans $P_{i}$ are defined by approximation of the points $M_{i}$ and containing the lines $D_{ij}$. We can draw the normal \vec{n_{i}} of each plan $P_{i}$ that include the point $M_{i}$.

We obtain a linear system with three vectors' equations and three unknown factors, after that is seeking to obtain the intersection point \textbf{CR$_{BT}$} using least squares minimization method. In reality, the directions of these three normal lines cross (nearly at r = 1.2~mm) in only one point noted \textbf{CR$_{BT}$} which corresponds with stiffness center.

\subsection{Determination of the attached plane of the tool tip displacements}
\label{sec:DEterminationDuPlanDesDEplacements}

The modeling of the attached plane of the tool tip displacements starting from the experimental results is carried out using the software Mathcad$^{\copyright}$. We seek to determine the plane that passes by the points cloud representing tool point displacements (Fig.~\ref{fig13}) :

\begin{eqnarray*}
	ax + by + cz + d = 0 .
\end{eqnarray*}

Using the parameter $e_{rr}$ for the errors, the following is obtained :

\begin{eqnarray*}
	e_{rr}(x, y,z, x_{p}, y_{p}, z_{p}, x_{n}, y_{n}, z_{n}) = \\
	\left[M(x, y, z) - P(x_{p},y_{p}, z_{p}).n_{u}(x_{n}, y_{n}, z_{n})\right],
\end{eqnarray*}

where:

\begin{eqnarray}
	M = \left\{x, y, z\right\}^{t}, P\left\{x_{p}, y_{p}, z_{p}\right\}^{t}, n_{u} = \left\{x_{n}, y_{n}, z_{n}\right\}^{t} ,
\end{eqnarray}

in this equation, superior index t indicates the transposition operation.
Let us express the $E_{rr}$ function using $e_{rr}$ and introduce the components of displacements $(u_{x_{i}}, u_{y_{i}}, u_{z_{i}})$ on the three space directions, it comes:

\begin{eqnarray*}
	E_{rr}(x, y,z, x_{p}, y_{p}, z_{p}, x_{n}, y_{n}, z_{n}) = \\ \sum^{N}_{i=0}e_{rr}(u_{x_{i}}, u_{y_{i}},u_{z_{i}}, x_{p}, y_{p}, z_{p}, x_{n}, y_{n}, z_{n}) ,
\end{eqnarray*}

and : 

\begin{eqnarray*}
	x_{p}=\frac{\sum^{N}_{i=0}u_{x_{i}}}{N}, y_{p}=\frac{\sum^{N}_{i=0}u_{y_{i}}}{N}, z_{p}=\frac{\sum^{N}_{i=0}u_{z_{i}}}{N} .
\end{eqnarray*}

The vector $\vec{V}$ is computed now by minimization:

\begin{eqnarray*}
	\vec{V} = min (E_{rr}, x_{p}, y_{p},z_{p}, x_{n}, y_{n}, z_{n}),
\end{eqnarray*}
 
where:

\begin{eqnarray*}
	\vec{V} = \left\{
	\begin{array}{c}
	V_{1} = x_{p}, \\
	V_{2} = y_{p},\\
	V_{3} = z_{p},\\
	V_{4} = x_{n},\\
	V_{5} = y_{n},\\
	V_{6} = z_{n}.
	
\end{array}\right.
\end{eqnarray*}

It results from it that the $n_{u}$ normal components in the plane of the tool tip displacements are given by:

\begin{eqnarray*}
	\vec{n_{u}} = \left\{
	\begin{array}{c}
	
	V_{4}, \\
	V_{5}, \\
	V_{6}.
	
\end{array}\right.
\end{eqnarray*}

For the present case study in the section~\ref{elldepl} (Table~\ref{tabl-3}), with ap =~5~mm and f =~0.1~mm, it comes:

\begin{eqnarray*}
	\vec{n_{u}} = \left\{
	\begin{array}{c}
	
	- 0.056 ,\\
	- 0.216 ,\\
	0.975 ,
	
\end{array}\right.
\end{eqnarray*}

the equation of the searched plane is then :

\begin{eqnarray*}
	P_{u}(s, t) = V_{p} + s.u_{1} + t.u_{2},
\end{eqnarray*}

where $s$ and $t$ are constants, and $\vec{V_{p}}$ is equal with:

\begin{eqnarray*}
	\vec{V_{p}} = \left\{
	\begin{array}{c}
	
	\vec{V_{1}} ,\\
	\vec{V_{2}} ,\\
	\vec{V_{3}} ,
	
\end{array}\right.
\end{eqnarray*}

and $\vec{u_{1}}$, $\vec{u_{2}}$ are given by:

\begin{eqnarray*}
	\vec{u_{1}} = \frac{\vec{u_{0} }\wedge \vec{n_{u}}}{\left\|\vec{n_{0}} \wedge \vec{u_{u}}\right\|},  \vec{u_{2}} = \vec{n }\wedge \vec{u_{1} } ,
\end{eqnarray*}

$\vec{u_{0}}$ being the plane orientation vector:

\begin{eqnarray*}
	\vec{u_{0}} = \left\{
	\begin{array}{c}
	
	1 ,\\
1 ,\\
	1 .
	
\end{array}\right.
\end{eqnarray*}

\subsection{Ellipse approximation}
\label{sec:ApproximationDeLEllipse}

Using the ellipse plane determination, it is possible to build the matrix $M_{rep}$ expressed by:

\begin{eqnarray*}
	\left[M_{rep}\right] = 
	\begin{array}{|ccc|}
	
	u_{1_{1}}& u_{2_{1}} & n_{u1} \\
	 u_{1_{2}}& u_{2_{2}} & n_{u1} \\
	u_{1_{2}}& u_{2_{3}} & n_{u1} 
	
\end{array} .
\end{eqnarray*}

Another matrix is defined by :  

\begin{eqnarray*}
	[ M_{u} ] = [ U ] - [ P ] ,
\end{eqnarray*}

with $[ P ]$ defined by the equation (2) and $[ U ]$, being the vector of the tool tip point displacement given by:

\begin{eqnarray*}
	\left[U\right] = \left\{
	\begin{array}{c}
	
	u_{x} \\
u_{y} \\
	u_{z} 
	
\end{array}\right\}	.
\end{eqnarray*}

Now, we determine the new ellipse configuration centered on the origin of the basic reference system (x, y, z):

\begin{eqnarray*}
	[ M_{u_{2}} ] = [ M_{rep} ]^{t} . [ M_{u} ] .
\end{eqnarray*}

The new displacements of the tool tip are :

\begin{eqnarray*}
	[ U_{1} ] = [ M_{u_{2}} ] .
\end{eqnarray*}

By changing the coordinate system, we project the ellipse in the new two-dimensional reference system (Fig.~\ref{fig21}), ($\alpha_{1},\alpha_{2}$) using the rotation matrix  $ [ M_{rot} ]$

\begin{eqnarray*}
	\left[M_{rot}\right] = 
	\begin{array}{|ccc|}
	
	\cos (\theta_{\alpha}) & - \sin (\theta_{\alpha}) & 0 \\
	 \sin (\theta_{\alpha}) & \cos (\theta_{\alpha}) & 0 \\
	0 & 0 & 1 
	
\end{array} .
\end{eqnarray*}

The angle $\theta_{\alpha}$ is obtained using minimization method and Mathcad$^{\copyright}$ software. We obtain:

\begin{figure}[htbp]
	\centering
		\includegraphics[width=0.50\textwidth]{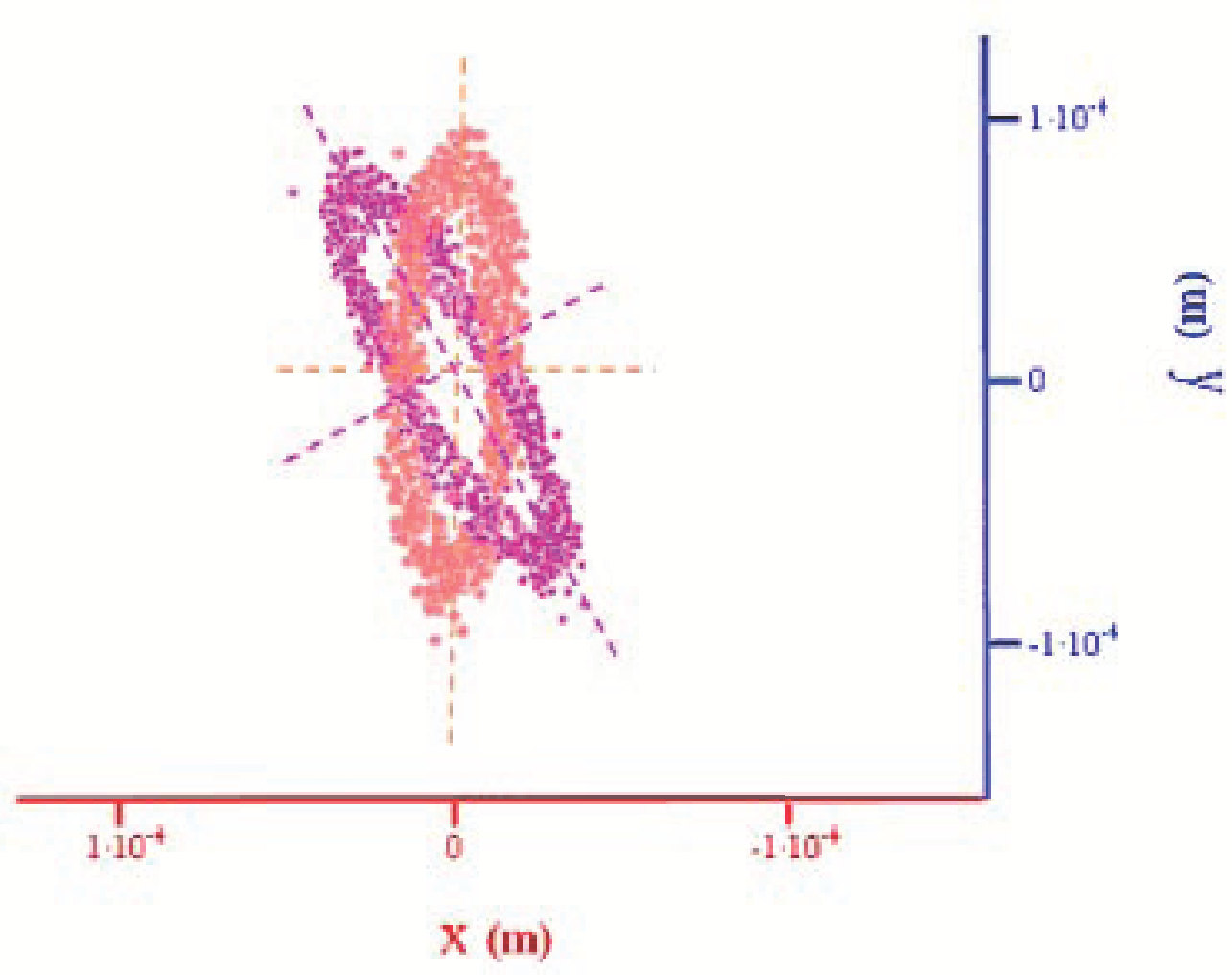}
	\caption{2D projected ellipse attached of the tool tip point displacements}
	\label{fig21}
\end{figure}

\begin{eqnarray*}
	[ M_{u_{3}} ] = [ M_{rot} ]^{t} . [ M_{u_{2}} ] ,
\end{eqnarray*}

and after that is obtained the tool tip displacement related to the base system ($\alpha_{1},\alpha_{2}$): 

\begin{eqnarray*}
	[ U_{2} ] = [ M_{u_{3}} ] .
\end{eqnarray*}

Consequently, the ellipse equation is:

\begin{eqnarray*}
\frac{(u_{\alpha_{1}})^{2}}{a^{2}} + \frac{(u_{\alpha_{2}})^{2}}{b^{2}} = 1 ,
\end{eqnarray*}

using parameters f =~0.1~mm/rev and ap =~5~mm. 
 
\begin{acknowledgements}
The authors would like to thank the \textit{Centre National de la Recherche Scientifique} (UMR 5469) for the financial support to accomplish the project.

\end{acknowledgements}

\end{document}